\documentclass[a4paper,11pt]{article}
\usepackage{bm,mathrsfs,amsmath,amsthm,amssymb,amscd,longtable,array,graphicx}
\newcommand{\nc}{\newcommand}
\nc{\rnc}{\renewcommand}
\nc{\nn}{\nonumber}
\nc{\der}{{\partial}}
\rnc{\Im}{{\rm{Im}\,}}
\rnc{\Re}{{\rm{Re}\,}}
\nc{\db}{\displaybreak[0]\\}
\nc{\bra}{\langle}
\nc{\ket}{\rangle}
\nc{\bs}{\boldsymbol}

\newtheorem{theorem}{Theorem}[section]

\newtheorem{proposition}[theorem]{Proposition}

\theoremstyle{definition}
\newtheorem{definition}[theorem]{Definition}

\numberwithin{equation}{section}

\numberwithin{equation}{section}

\begin{document}%
%
\title{Scalar products of the elliptic Felderhof model \\
and elliptic Cauchy formula}

\author{
Kohei Motegi \thanks{E-mail: kmoteg0@kaiyodai.ac.jp}
\\\\
{\it Faculty of Marine Technology, Tokyo University of Marine Science and Technology,}\\
 {\it Etchujima 2-1-6, Koto-Ku, Tokyo, 135-8533, Japan} \\
\\\\
\\
}

\date{}

\maketitle

\begin{abstract}
We analyze the scalar products of the elliptic Felderhof model
introduced by Foda-Wheeler-Zuparic as an elliptic extension
of the trigonometric face-type Felderhof model by Deguchi-Akutsu.
We derive the determinant formula for the scalar products
by applying the Izergin-Korepin technique developed by Wheeler
to investigate the scalar products of integrable lattice models.
By combining the determinant formula for the scalar products with the
recently-developed Izergin-Korepin technique to analyze the wavefunctions,
we derive a Cauchy formula for elliptic Schur functions.
\end{abstract}

Elliptic integrable models are classes of integrable models
described by elliptic functions.
Investigations of elliptic integrable models
lead to new discoveries of mathematical structures. An instance is
the notion of elliptic quantum groups
\cite{Felder,FV,Konno1,JKOS}
which are extensions of the quantum groups \cite{FST,Dr,J},
introduced through the analysis of the eight-vertex model,
eight-vertex solid-on-solid model and their generalizations
\cite{Baxter,eight,ABF,JMO,DJKMO}.
Recently, there are also progresses on partition functions of
the eight-vertex solid-on-solid models
\cite{
PRS,Ros,YZ,FiKi,Chinesegroup1,Chinesegroup2,Galleas1,Galleas2,GL,Lamers,Konno2,Bor,FRV,RTV}
from the viewpoint of the quantum inverse scattering method \cite{KBI,Re},
vertex operator method \cite{JM} and so on.

In this paper, we investigate another class of elliptic integrable model.
We analyze the scalar products of the elliptic Felderhof model
introduced by Foda-Wheeler-Zuparic \cite{FWZ}
as an extension of the face-type Felderhof model \cite{Felderhof} by Deguchi-Akutsu \cite{DA}.
The elliptic Felderhof model (Foda-Wheeler-Zuparic model),
and its closely related
elliptic Perk-Schultz model (Okado-Deguchi-Fujii-Martin model) constructed by
Okado \cite{Okado}, Deguchi-Fujii \cite{DF} and Deguchi-Martin \cite{DM}
as an elliptic extension of the Perk-Schultz model
\cite{PS}, are interesting models to be investigated,
since the corresponding trigonometric models were discovered
by number theorists recently
to be related with automorphic representation theory
and deformations of Weyl character formulas (Tokuyama formulas)
for symmetric functions. Bump-Brubaker-Friedberg \cite{BBF}
constructed free-fermion models by themselves and
showed that the wavefunctions are given as a
product of a deformed Vandermonde determinant
and Schur functions.
One of the consequences of their results
is the natural construction
of the Tokuyama formula \cite{To} as wavefunctions
of integrable models, which is a one-parameter deformation
of the Weyl character formula for the Schur functions.
Their result is the one of the main motivations
to study the elliptic Felderhof model of Foda-Wheeler-Zuparic
and the elliptic Perk-Schultz model of Okado, Deguchi-Fujii and Deguchi-Martin,
since these models can be regarded as
elliptic analogues of the free-fermion model which Bump-Brubaker-Friedberg
introduced and analyzed
(the quantum group structure of the
trigonometric models can be found in \cite{DA,Yamane,Murakami} for example)
. There are not so much studies on the partition functions
of these elliptic models. Foda-Wheeler-Zuparic
showed the factorization of the domain wall boundary partition functions
of these models \cite{FWZ} by applying the Izergin-Korepin technique
\cite{Ko,Iz},
which is a classical
method to analyze the domain wall boundary partition functions
of integrable models.
Recently, we extended the Izergin-Korepin technique
to be able to anlayze the wavefunctions \cite{MoIK,Moelliptic,Motr},
and showed that the wavefunctions of these elliptic models
are given as a deformed elliptic Vandermonde determinant
and elliptic symmetric functions which can be viewed
as elliptic Schur functions
(see Schlosser \cite{Sch} or Noumi \cite{Noumi,Noumi2}
for other types of elliptic Schur functions introduced
from the viewpoint of combinatorics, special functions
and classical integrable systems).
The results can be viewed as elliptic analogues
of the one by Bump-Brubaker-Friedberg.

In this paper, we investigate another special class
of partition functions called the scalar products.
One of the motivations to study this class of partition functions
comes from the recent active line of researches on
the application of the correspondence
between symmetric functions
and wavefunctions of integrable models
to derivations of various algebraic identities.
For the free-fermionic models, see \cite{OkTo,HK1,HK2,Iv,BBCG,Tabony,
BMN,HK,BS,BBB,LMP,ZZ,FCWZ,
ZYZ,Wheelerthesis,Zuparicthesis} for examples on integrability approach
to symmteric functions, as well as closely related non-intersecting
lattice paths approach.
There are also investigations on the
six-vertex models and face models related to the
XXX, XXZ and XYZ quantum integrable spin chains and $q$-boson models,
where the Schur, Grothendieck, Hall-Littlewood polynomials and their
generalizations appear as the wavefunctions.
See \cite{Bogo,BWZ,BW,WZnew,vDE,MS,Motegi,MS2,KS,Korff,Borodin,BP1,
TakeyamaHecke,Takeyama} for examples on various studies of these models.
Among these active studies, it was realized that the analysis
on the scalar products lead us to Cauchy formulas for symmetric functions.
Directly evaluating the scalar products to get determinant formulas
in one way, and comparing the expressions with
another way of evaluation by inserting completeness relation
and express it as the sum of products of the wavefunctions
whose explicit forms are given by symmetric functions,
one can get Cauchy formulas for symmetric functions.
This quantum integrability approach often enables us to derive
algebraic identities which are almost impossible to find by any other means.
In this paper, we apply this idea to the elliptic Felderhof model,
and derive the Cauchy formula for the elliptic Schur functions.
The main part of this paper is the direct evaluation
of the scalar products. We apply the Izergin-Korepin technique
developed by Wheeler \cite{Wheeler} to derive the scalar products
of integrable models. In his paper, Wheeler showed that his
technique can be applied to the $U_q(sl_2)$ six-vertex model
to derive the Slavnov's determinant formula \cite{Slavnov}
for the XXZ spin chain for example,
by introducing and listing the properties
which uniquely defines the intermediate scalar products,
and showing the explicit determinant forms satisfying all the
properties.
We apply his technique to the elliptic Felderhof model and
obtain the determinant formula for the scalar products.
Together with our results on the correspondence between the wavefunctions
and the elliptic Schur functions \cite{Moelliptic,Motr}
obtained by the Izergin-Korepin analysis on the wavefunctions,
we derive the Cauchy formula for the elliptic Schur functions.

This paper is organized as follows.
In Section 2, we recall the properties
of theta functions and the Foda-Wheeler-Zuparic
(elliptic Felderhof) model.
In Section 3, we introduce the scalar products,
and derive the determinant formula by applying the
Izergin-Korepin technique developed by Wheeler.
In Section 4, by combining with another evaluation
of the scalar products using the correspondence
between the wavefunctions and the elliptic Schur functions,
we derive the Cauchy formula
for the elliptic Schur functions.
Section 5 is devoted to the conclusion of this paper.

\section{Foda-Wheeler-Zuparic (elliptic Felderhof) model}

In this section, we first introduce elliptic functions
and list their properties,
and introduce the Foda-Wheeler-Zuparic model
which is an elliptic analogue of the Felderhof model.
The theta functions $H(u)$ is
\begin{align}
H(u)=2 \mathrm{sinh} u
\prod_{n=1}^\infty (1-2{\bf q}^{2n} \mathrm{cosh}(2u)+{\bf q}^{4n})(1-{\bf q}^{2n}),
\end{align}
where ${\bf q}$ is the elliptic nome $(0 < {\bf q} < 1)$.
For the description of the matrix elements of the dynamical $R$-matrix
of the elliptic Felderhof model,
we introduce the following notation
\begin{align}
[u]=H(\pi i u).
\end{align}

The theta function $[u]$ is an odd function $[-u]=-[u]$
and satisfies the quasi-periodicities
\begin{align}
[u+1]&=-[u], \label{qpone} \\
[u-i \mathrm{log}({\bf q})/\pi]&=-{\bf q}^{-1}
\mathrm{exp} (-2 \pi i u) [u] \label{qptwo}.
\end{align}

We use the following property about the elliptic polynomials \cite{PRS,FS}
presented below.

A character is a group homomorphism
$\chi$ from multiplicative groups
$\Gamma=\mathbf{Z}+\tau \mathbf{Z}$ to $\mathbf{C}^\times$.
An $N$-dimensional space $\Theta_N(\chi)$
is defined for each character $\chi$ and positive integer $N$,
which consists of holomorphic functions $\phi(y)$ on $\mathbf{C}$
satisfying the quasi-periodicities
\begin{align}
\phi(y+1)&=\chi(1) \phi(y), \label{propertyuseone} \\
\phi(y+\tau)&=\chi(\tau) e^{-2 \pi i Ny-\pi i N \tau}\phi(y).
\label{propertyusetwo}
\end{align}
The elements of the space $\Theta_N(\chi)$ are called elliptic polynomials.
The space $\Theta_N(\chi)$ is $N$-dimensional \cite{PRS,FS}
and the following fact holds for the elliptic polynomials.
\begin{proposition} \cite{PRS,FS} \label{propositionelliptic}
Suppose there are two elliptic polynomials $P(y)$ and $Q(y)$
in $\Theta_N(\chi)$, where $\chi(1)=(-1)^N$, $\chi(\tau)=(-1)^N e^\alpha$.
If those two polynomials are equal $P(y_j)=Q(y_j)$
at $N$ points $y_j$, $j=1,\dots,N$ satisfying
$y_j-y_k \not\in \Gamma$, $\sum_{k=1}^N y_k-\alpha \not\in \Gamma$,
then the two polynomials are exactly the same $P(y)=Q(y)$.
\end{proposition}
This property ensure the uniqueness of the Izergin-Korepin analysis
on the wavefunctions of elliptic integrable models.
For example, it is used in \cite{PRS,Ros,YZ} on the analysis on
the domain wall boundary partition functions of the
eight-vertex solid-on-solid model \cite{ABF}.
Note that the property above is an elliptic analogue of 
the following fact for ordinary polynomials:
if $P(y)$ and $Q(y)$ are polynomials of degree $N-1$ in $y$,
and if these polynomials match at $N$ distinct points,
then the two polynomials are exactly the same.

\begin{figure}[ht]
\includegraphics[width=15cm]{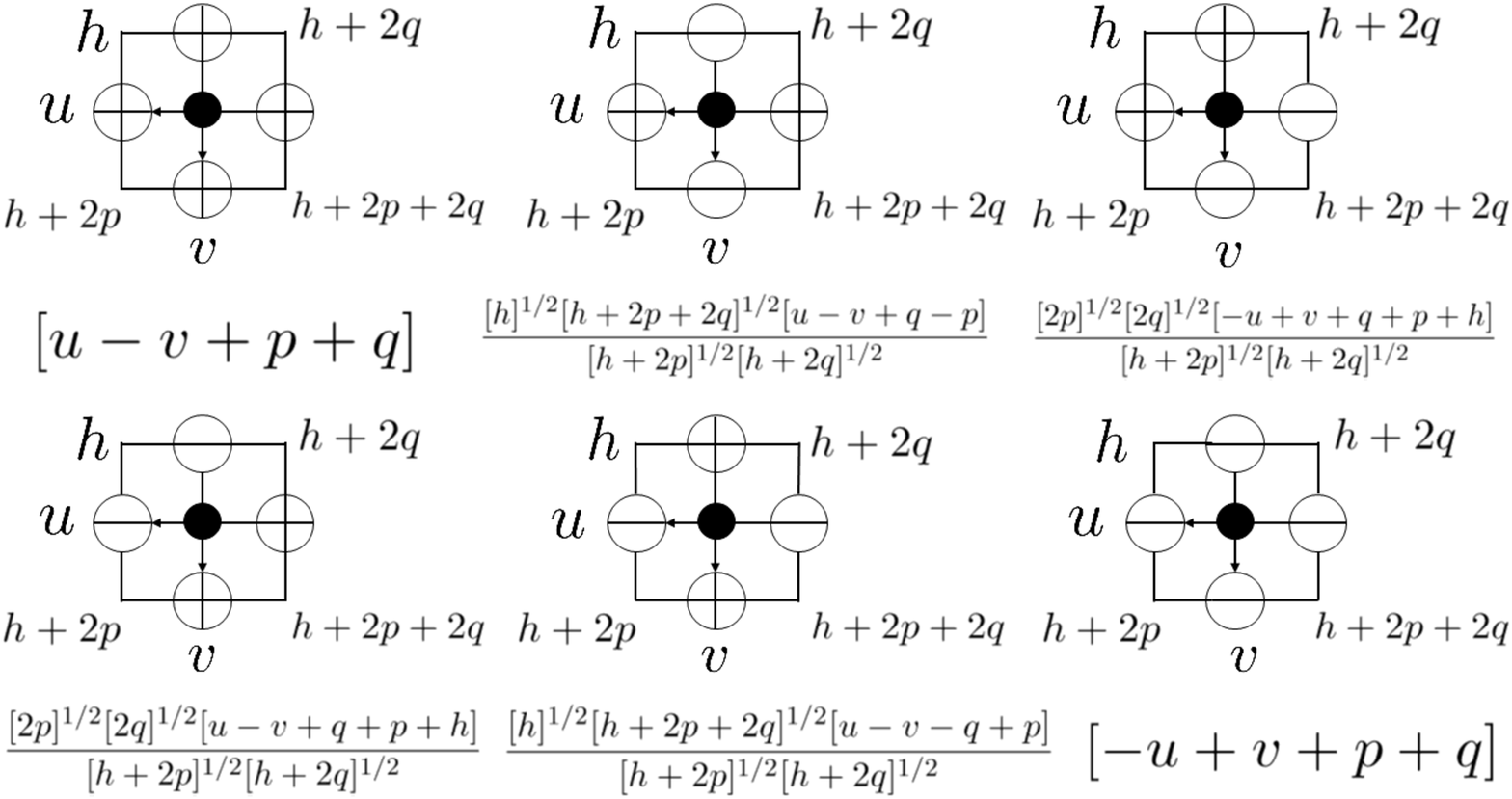}
\caption{The matrix elements of the elliptic Felderhof model
\protect\eqref{rmatrix}.
The states $|0 \rangle$, $\langle 0|$ are represented
as $\oplus$, while the states $|1 \rangle$, $\langle 1|$
are represented as $\ominus$.
This kind of graphical representation for the case
of trigonometric vertex models can be found in
Bump-Brubaker-Friedberg \cite{BBF}
and Bump-McNamara-Nakasuji \cite{BMN}
for example.
}
\label{pictureloperator}
\end{figure}

The trigonometric face-type Felderhof model was first introduced by
Deguchi-Akutsu \cite{DA},
and its elliptic extension was constructed by Foda-Wheeler-Zuparic
\cite{FWZ}.
The dynamical $R$-matrix of the elliptic Felderhof model
is given by \cite{FWZ}
\begin{align}
&R_{ab}(u,v|p,q|h) \nonumber \\
=&\left( 
\begin{array}{cccc}
[u-v+p+q] & 0 & 0 & 0 \\
0 & \frac{[h]^{1/2}[h+2p+2q]^{1/2}[u-v+q-p]}{[h+2p]^{1/2}[h+2q]^{1/2}} &
\frac{[2p]^{1/2}[2q]^{1/2}[-u+v+q+p+h]}{[h+2p]^{1/2}[h+2q]^{1/2}}  & 0 \\
0 & \frac{[2p]^{1/2}[2q]^{1/2}[u-v+q+p+h]}{[h+2p]^{1/2}[h+2q]^{1/2}} & 
\frac{[h]^{1/2}[h+2p+2q]^{1/2}[u-v-q+p]}{[h+2p]^{1/2}[h+2q]^{1/2}} & 0 \\
0 & 0 & 0 & [-u+v+p+q]
\end{array}
\right), \label{rmatrix}
\end{align}
acting on the tensor product $W_a \otimes W_b$
of the complex two-dimensional space $W_a$.
The parameters $u$ and $v$ are spectral parameters,
and $p$ and $q$ are complex parameters.
$h$ is called as the height or dynamical variable.
One can think that the space $W_a$ carries the parameters
$u$ and $p$, while the
parameters $v$ and $q$ are associated with the space $W_b$.
See Figure \eqref{pictureloperator}
for the graphical representations of the dynamical $R$-matrix \eqref{rmatrix}.

We denote the orthonormal basis of $W_a$ and its dual as
$\{|0 \rangle_a, |1 \rangle_a \}$ and $\{{}_a \langle 0|, {}_a \langle 1|\}$,
and the matrix elements of the dynamical $R$-matrix as
$
{}_a \langle \gamma | {}_b \langle \delta | R_{a b}(u,v|p,q|h)
|\alpha \rangle_a | \beta \rangle_b=[R(u,v|p,q|h)]_{\alpha \beta}^{\gamma \delta}
$. The matrix elements of the dynamical $R$-matrix are explicitly given as
\begin{align}
{}_a \langle 0| {}_b \langle 0 | R_{a b}(u.v|p,q|h)
|0 \rangle_a | 0 \rangle_b&=[u-v+p+q], \\
{}_a \langle 0| {}_b \langle 1 | R_{a b}(u,v|p,q|h)
|0 \rangle_a | 1 \rangle_b&=\frac{[h]^{1/2}[h+2p+2q]^{1/2}[u-v+q-p]}{[h+2p]^{1/2}[h+2q]^{1/2}}, \\
{}_a \langle 0| {}_b \langle 1 | R_{a b}(u,v|p,q|h)
|1 \rangle_a | 0 \rangle_b&=\frac{[2p]^{1/2}[2q]^{1/2}[-u+v+q+p+h]}{[h+2p]^{1/2}[h+2q]^{1/2}}, \\
{}_a \langle 1| {}_b \langle 0 | R_{a b}(u,v|p,q|h)
|0 \rangle_a |1 \rangle_b&=\frac{[2p]^{1/2}[2q]^{1/2}[u-v+q+p+h]}{[h+2p]^{1/2}[h+2q]^{1/2}}, \\
{}_a \langle 1| {}_b \langle 0 | R_{a b}(u,v|p,q|h)
|1 \rangle_a | 0 \rangle_b&=\frac{[h]^{1/2}[h+2p+2q]^{1/2}[u-v-q+p]}{[h+2p]^{1/2}[h+2q]^{1/2}}, \\
{}_a \langle 1| {}_b \langle 1 | R_{a b}(u,v|p,q|h)
|1 \rangle_a | 1 \rangle_b&=[-u+v+p+q].
\end{align}

In statistical physics,
$| 0 \rangle$ or its dual $\langle 0|$
can be regarded as a hole state,
while $| 1 \rangle$ or its dual $\langle 1|$
can be interpretted as a particle state.
We thus sometimes use the terms hole states and particle states
to describe states constructed from
$| 0 \rangle$, $\langle 0|$, $| 1 \rangle$ and $\langle 1|$
since they are convenient for the description of the states.

For later convenience,
we also define the following Pauli spin operators
$\sigma^+$ and $\sigma^-$ as operators acting on the (dual) orthonomal
basis as
\begin{align}
&\sigma^+|1 \rangle=|0 \rangle, \ 
\sigma^+|0 \rangle=0, \ 
\langle 0|\sigma^+=\langle 1|, \
\langle 1|\sigma^+=0, 
\\
&\sigma^-|0 \rangle=|1 \rangle, \
\sigma^-|1 \rangle=0, \
\langle 1|\sigma^-=\langle 0|, \
\langle 0|\sigma^-=0.
\end{align}

\begin{figure}[ht]
\includegraphics[width=15cm]{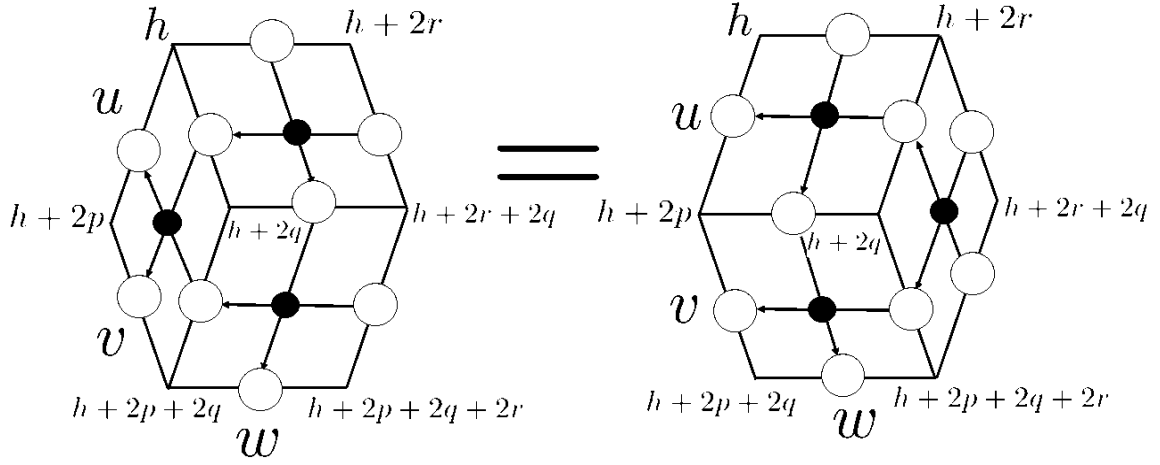}
\caption{The dynamical Yang-Baxter relation \protect\eqref{yangbaxter}.
The left and right figure represents
$R_{ab}(u,v|p,q|h)R_{ac}(u,w|p,r|h+2q)R_{bc}(v,w|q,r|h)$
and
$R_{bc}(v,w|q,r|h+2p)R_{ac}(u,w|p,r|h)R_{ab}(u,v|p,q|h+2r)$,
respectively.
}
\label{pictureyangbaxter}
\end{figure}

The dynamical $R$-matrix
\eqref{rmatrix}
satisfies the dynamical Yang-Baxter
(face-type Yang-Baxter, star-triangle)
relation (Figure \ref{pictureyangbaxter})

\begin{align}
&R_{ab}(u,v|p,q|h)R_{ac}(u,w|p,r|h+2q)R_{bc}(v,w|q,r|h) \nonumber \\
=&R_{bc}(v,w|q,r|h+2p)R_{ac}(u,w|p,r|h)R_{ab}(u,v|p,q|h+2r),
\label{yangbaxter}
\end{align}
acting on $W_a \otimes W_b \otimes W_c$.

To construct partition functions of integrable lattice models,
we identify one of the complex two-dimensional spaces
$W_b$
of the tensor product space $W_a \otimes W_b$
with the quantum space.
Let us denote the quantum space by $\mathcal{F}_j$,
and define the $L$-operator $L_{aj}(u,v|p,q|h)$ acting on
$W_a \otimes \mathcal{F}_j$ as 
\begin{align}
L_{aj}(u,v|p,q|h)
=R_{aj}(u,v|p,q|h).
\end{align}

The next step is to define the monodromy matrix from the $L$-operators.
For convenience, one denotes the sum of complex numbers
$q_1,q_2,\dots,q_j$ as $\overline{q_j}$
\begin{align}
\overline{q_j}:=\sum_{k=1}^j q_k.
\end{align}

The monodromy matrix $T_a(u|v_1,\dots,v_M|p|q_1,\dots,q_M|h)$
is the product of $L$-operators
\begin{align}
&T_a(u|v_1,\dots,v_M|p|q_1,\dots,q_M|h) \nonumber \\
=&L_{a1}(u,v_1|p,q_1|h)L_{a2}(u,v_2|p,q_2|h+2\overline{q_1}) \cdots
L_{aM}(u,v_M|p,q_M|h+2\overline{q_{M-1}}), \label{monodromy}
\end{align}
acting on $W_a \otimes \mathcal{F}_1 \otimes \cdots \otimes \mathcal{F}_M$.

\begin{figure}[ht]
\includegraphics[width=12cm]{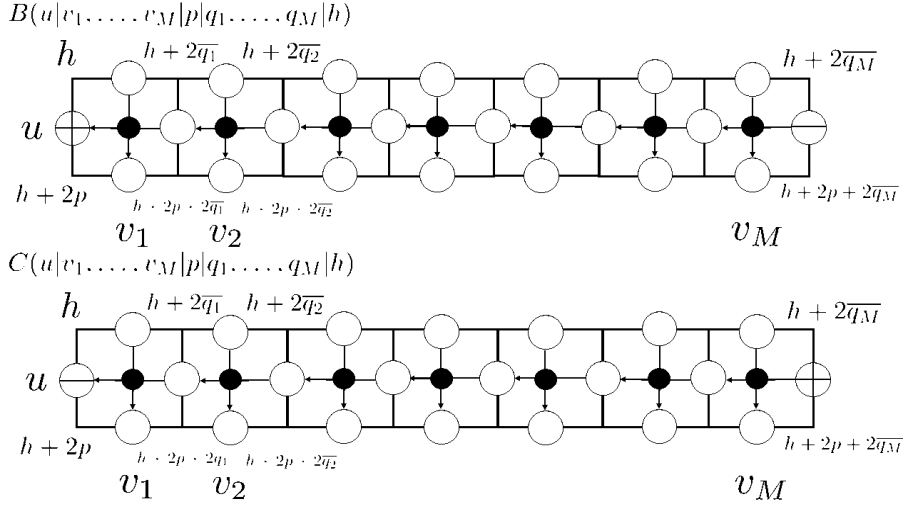}
\caption{The $B$-operator
$B(u|v_1,\dots,v_M|p|q_1,\dots,q_M|h)$
\protect\eqref{Boperator} (top)
and
the $C$-operator
$C(u|v_1,\dots,v_M|p|q_1,\dots,q_M|h)$
\protect\eqref{Coperator} (bottom).
}
\label{pictureprodcutofweights}
\end{figure}

The $B$-operator and the $C$-operator
are matrix elements of the monodromy matrix
\eqref{monodromy} with respect to the auxiliary space $W_a$
\begin{align}
B(u|v_1,\dots,v_M|p|q_1,\dots,q_M|h)
&={}_{a} \langle 0|T_a(u|v_1,\dots,v_M|p|q_1,\dots,q_M|h)
|1 \rangle_a, \label{Boperator} \\
C(u|v_1,\dots,v_M|p|q_1,\dots,q_M|h)
&={}_{a} \langle 1|T_a(u|v_1,\dots,v_M|p|q_1,\dots,q_M|h)
|0 \rangle_a, \label{Coperator}
\end{align}
which acts on $\mathcal{F}_1 \otimes \cdots \otimes \mathcal{F}_M$
and it dual $\mathcal{F}_1^* \otimes \cdots \otimes \mathcal{F}_M^*$
(Figure \ref{pictureprodcutofweights}).

\begin{figure}[ht]
\includegraphics[width=12cm]{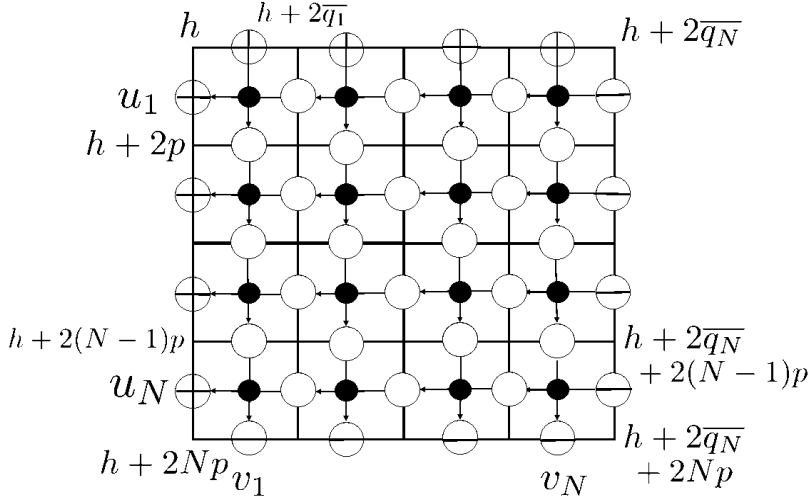}
\caption{The domain wall boundary partition functions
$Z_N(u_1,\dots,u_N|v_1,\dots,v_N|h)$ \protect\eqref{domainwall}.
}
\label{picturedomainwall}
\end{figure}

\begin{figure}[ht]
\includegraphics[width=15cm]{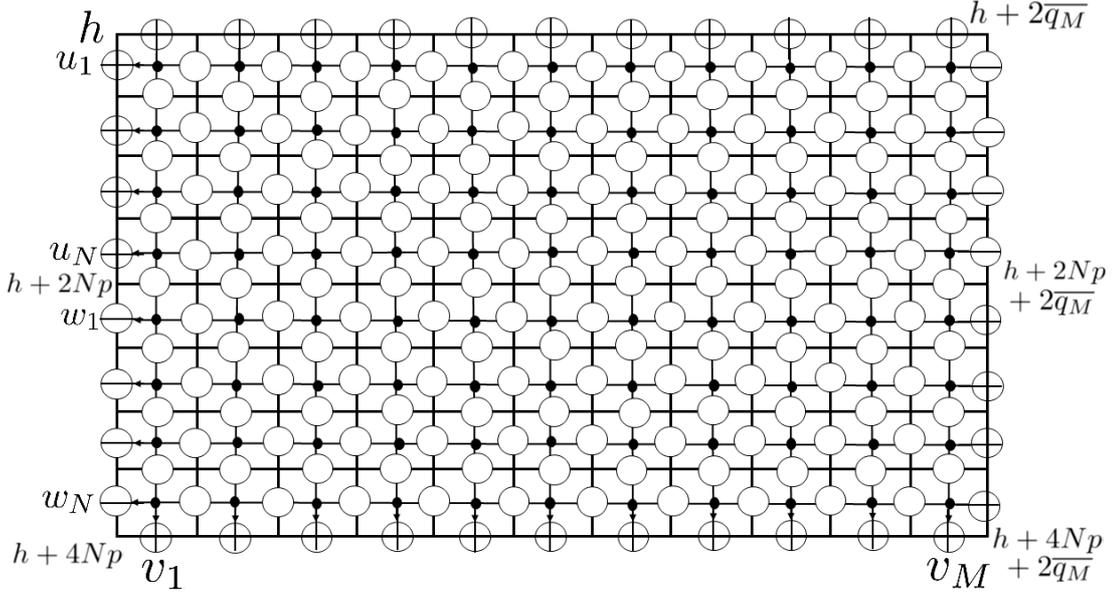}
\caption{The scalar products $Q_{M,N}(u_1,\dots,u_N|w_1,\dots,w_N|v_1,\dots,v_{M}|h)$
\protect\eqref{scalarproducts}.}
\label{ordinarypicturescalarproducts}
\end{figure}

\begin{figure}[ht]
\includegraphics[width=15cm]{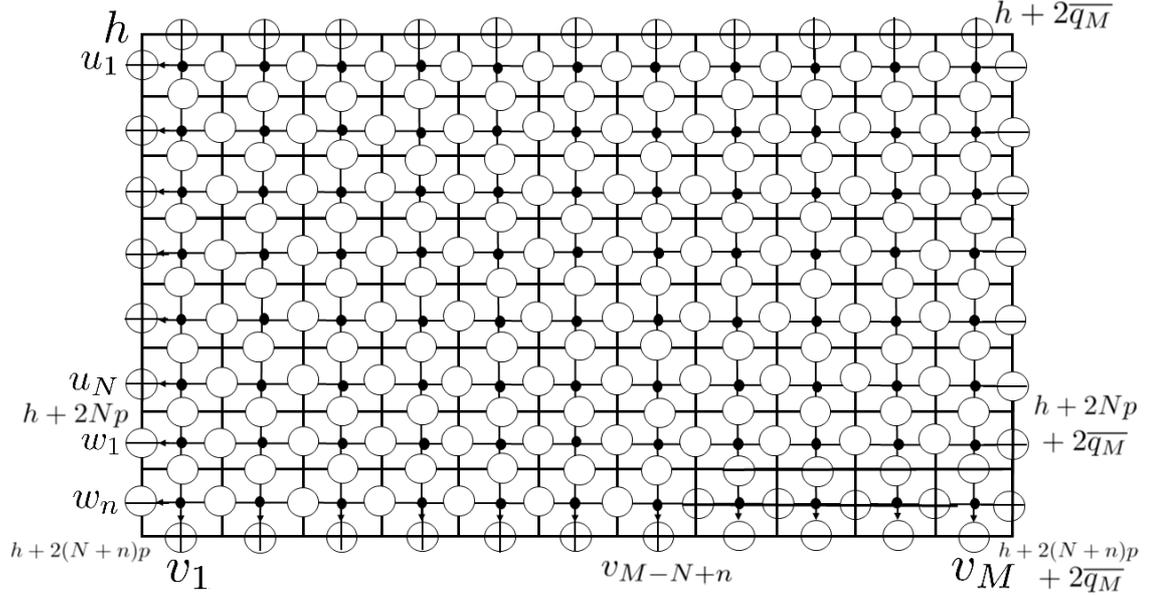}
\caption{The intermediate scalar products $Q_{M,N,n}(u_1,\dots,u_N|w_1,\dots,w_n|v_1,\dots,v_{M}|h)$
\protect\eqref{scalarproducts}. One notes that the
dynamical $R$-matrices in the right part of the bottom row
are already frozen due to the ice-rule of the dynamical $R$-matrix.
}
\label{ordinarypictureintermediatescalarproducts}
\end{figure}

\begin{figure}[ht]
\includegraphics[width=15cm]{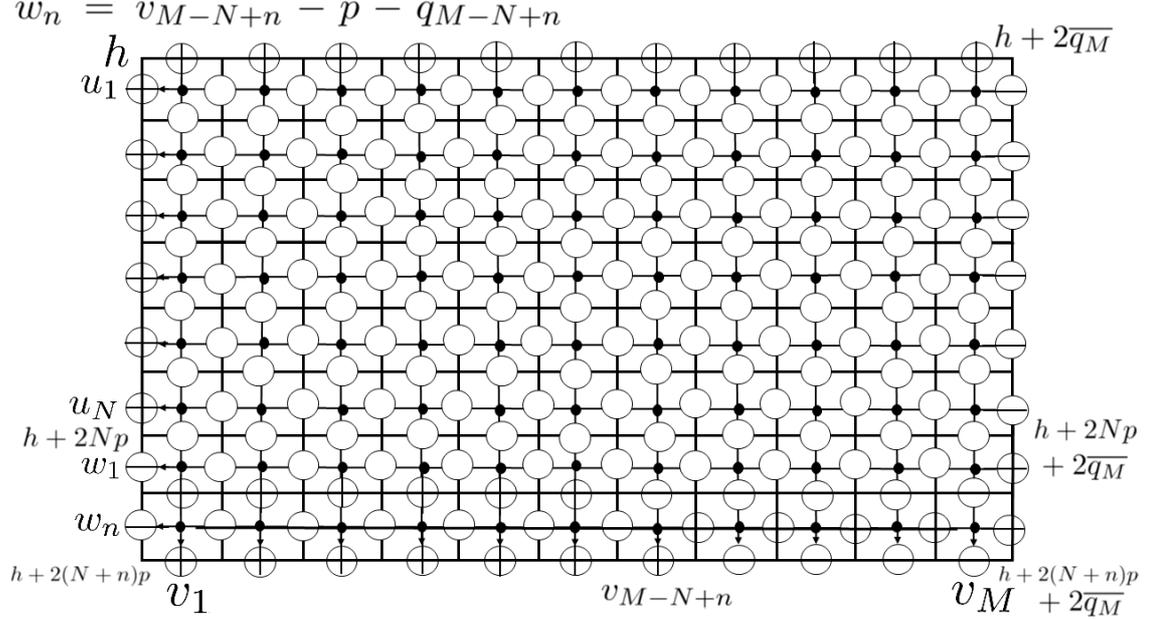}
\caption{The intermediate scalar products $Q_{M,N,n}(u_1,\dots,u_N|w_1,\dots,w_n|v_1,\dots,v_{M}|h)$
\protect\eqref{scalarproducts} evaluated at $w_n=v_{M-N+n}-p-q_{M-N+n}$,
which gives \protect\eqref{specialpointofintermediatescalarproducts}
since the dynamical $R$-matrices at the bottom row
are frozen and the remaining part is
$Q_{M,N,n-1}(u_1,\dots,u_N|w_1,\dots,w_{n-1}|v_1,\dots,v_{M}|h)$.
}
\label{ordinarypictureintermediatescalarproductsspecialpoint}
\end{figure}

The following domain wall boundary partition functions
(Figure \ref{picturedomainwall})
is one of the most well-investigated classes of partition functions \cite{Ko}
\begin{align}
&Z_N(u_1,\dots,u_N|v_1,\dots,v_N|h)
={}_N \langle \overline{\Omega}|
B(u_N|v_1,\dots,v_M|p|q_1,\dots,q_M|h+2(N-1)p) \nonumber \\
&\times \cdots \times
B(u_2|v_1,\dots,v_M|p|q_1,\dots,q_M|h+2p)
B(u_1|v_1,\dots,v_M|p|q_1,\dots,q_M|h)|\Omega \rangle_N.
\label{domainwall}
\end{align}
Here,
$|\Omega \rangle_N:=|0 \rangle_1 \otimes \cdots \otimes |0 \rangle_N
\in \mathcal{F}_1 \otimes \cdots \otimes \mathcal{F}_N$
and
${}_{N} \langle \overline{\Omega} |:={}_1 \langle 1 | \otimes \cdots \otimes
{}_N \langle 1 |
\in \mathcal{F}_1^* \otimes \cdots \otimes \mathcal{F}_N^*$
are the vacuum state and the dual particle-occupied state
in the tensor product of quantum spaces.

In the paper in which the elliptic Felderhof model was introduced,
Foda-Wheeler-Zuparic
showed the following factorized expression
for the domain wall boundary partition functions
\cite{FWZ,Wheelerthesis,Zuparicthesis}.

\begin{theorem} (Foda-Wheeler-Zuparic \cite{FWZ}) \label{theoremFWZ}
The domain wall boundary partition functions
of the elliptic Felderhof model
$Z_N(u_1,\dots,u_N|v_1,\dots,v_N|h)$
has the following factorized form
\begin{align}
Z_N(u_1,\dots,u_N|v_1,\dots,v_N|h)
=&\frac{[h+\sum_{j=1}^N v_j-\sum_{j=1}^N u_j+Np+\sum_{j=1}^N q_j]}
{[h+2 \sum_{j=1}^N q_j]^{1/2} [h+2Np]^{1/2}} \nonumber \\
&\times
\prod_{j=1}^N [2p]^{1/2} [2q_j]^{1/2}
\prod_{1 \le j < k \le N}[u_j-u_k+2p][v_k-v_j+q_j+q_k]. \label{expressionFWZ}
\end{align}
\end{theorem}
We use \eqref{expressionFWZ} for the analysis on the
scalar products of the elliptic Felderhof model
in the next section.

\section{Scalar Products}\label{sec3}

In this section, we introduce the scalar products
of the Foda-Wheeler-Zuparic model, and prove the determinant formula.
The scalar products are defined as the following partition functions
(Figure \ref{ordinarypicturescalarproducts})
\begin{align}
&Q_{M,N}(u_1,\dots,u_N|w_1,\dots,w_N|v_1,\dots,v_{M}|h)
\nonumber \\
=&{}_{M} \langle \Omega|
C(w_N|v_1,\dots,v_{M}|p|q_1,\dots,q_M|h+2(2N-1)p) \nonumber \\
&\times \cdots \times
C(w_1|v_1,\dots,v_{M}|p|q_1,\dots,q_M|h+2Np) \nonumber \\
&\times B(u_N|v_1,\dots,v_{M}|p|q_1,\dots,q_M|h+2(N-1)p) \nonumber \\
&\times \cdots \times
B(u_1|v_1,\dots,v_{M}|p|q_1,\dots,q_M|h)|\Omega \rangle_{M},
\label{scalarproducts}
\end{align}
where
$|\Omega \rangle_M:=|0 \rangle_1 \otimes \cdots \otimes |0 \rangle_M
\in \mathcal{F}_1 \otimes \cdots \otimes \mathcal{F}_M$
and
${}_{M} \langle \Omega |:={}_1 \langle 0 | \otimes \cdots \otimes
{}_M \langle 0 |
\in \mathcal{F}_1^* \otimes \cdots \otimes \mathcal{F}_M^*$
are the vacuum state and the dual vacuum state
in the tensor product of quantum spaces.

The main result of this paper is the following
determinant formula for the scalar products of the elliptic Felderhof model.

\begin{theorem} \label{maintheoremscalarproducts}
We have the following determinant formula for the
scalar products of the elliptic Felderhof model
\begin{align}
&Q_{M,N}(u_1,\dots,u_N|w_1,\dots,w_N|v_1,\dots,v_{M}|h) \nonumber \\
=&[2p]^N \frac{[h+2Np]^{1/2}[h+2Np+2\overline{q_M}]^{1/2}}
{[h+4Np]^{1/2} [h+2\overline{q_M}]^{1/2}}
\prod_{1 \le j < k \le N}
\frac{[u_j-u_k+2p][w_j-w_k-2p]}{[u_j-u_k][w_j-w_k]} \nonumber \\
&\times \mathrm{det}_N
\Bigg(
\frac{\frac{[w_j+h+2Np-u_k]}{[h+2Np]}a(w_j)d(u_k)
-\frac{[w_j+h+2Np-u_k+2\overline{q_M}]}{[h+2Np+2\overline{q_M}]}a(u_k)d(w_j)
}{[u_k-w_j]}
\Bigg), \label{determinantscalarproducts}
\end{align}
where $\overline{q_j}=\sum_{k=1}^j q_k$,
and $a(u)$ and $d(u)$ are given by
\begin{align}
a(u)=\prod_{\ell=1}^M [u-v_\ell+p+q_\ell], \ \ \
d(u)=\prod_{\ell=1}^M [u-v_\ell+p-q_\ell].
\end{align}
\end{theorem}

We apply Wheeler's method \cite{Wheeler}
which extends the Izergin-Korpein technique
\cite{Iz,Ko} from the domain wall boundary partition functions
to the scalar products.
To this end, we introduce the following intermediate scalar products
\cite{Wheeler} which is an intermediate object
between the scalar products and the domain wall boundary partition functions
(Figure \ref{ordinarypictureintermediatescalarproducts})

\begin{align}
&Q_{M,N,n}(u_1,\dots,u_N|w_1,\dots,w_n|v_1,\dots,v_{M}|h) \nonumber \\
=&\langle 0^{M-N+n} 1^{N-n}|
C(w_n|v_1,\dots,v_{M}|p|q_1,\dots,q_M|h+2(N+n-1)p) \nonumber \\
&\times \cdots \times
C(w_1|v_1,\dots,v_{M}|p|q_1,\dots,q_M|h+2Np) \nonumber \\
&\times B(u_N|v_1,\dots,v_{M}|p|q_1,\dots,q_M|h+2(N-1)p) \nonumber \\
&\times \cdots \times
B(u_1|v_1,\dots,v_{M}|p|q_1,\dots,q_M|h)|\Omega \rangle_{M},
\label{intermediatescalarproducts}
\end{align}
where
\begin{align}
\langle 0^{M-N+n} 1^{N-n}|
={}_1 \langle 0| \otimes \cdots \otimes {}_{M-N+n} \langle 0|
\otimes {}_{M-N+n+1} \langle 1| \otimes \cdots \otimes
{}_M \langle 1|.
\end{align}

The special case $n=N$ of the intermediate scalar products
corresponds to the scalar products
$Q_{M,N}(u_1,\dots,u_N|w_1,\dots,w_N|v_1,\dots,v_{M}|h)$,
while the case $n=0$ is essentially the domain wall boundary partition functions.

The first thing to do is to list the properties of the intermediate
scalar products which uniquely characterize it,
which is given below.

\begin{proposition} 
\label{aboutgeneralizedscalarproducts}
The intermediate scalar products of the elliptic Felderhof model \\
$Q_{M,N,n}(u_1,\dots,u_N|w_1,\dots,w_n|v_1,\dots,v_{M}|h)$
satisfies the following properties. \\
\\
 (1)
$\displaystyle \prod_{j=M-N+n+1}^M [w_n-v_j+q_j-p]^{-1}
Q_{M,N,n}(u_1,\dots,u_N|w_1,\dots,w_n|v_1,\dots,v_{M}|h)$
is an elliptic polynomial of $w_n$ in $\Theta_{M-N+n}(\chi)$.
\\
\\
 (2) The
intermediate scalar products
$Q_{M,N,n}(u_1,\dots,u_N|w_1,\dots,w_n|v_1,\dots,v_{M}|h)$
is invariant under the simultaneous exchange of
$v_j$, $q_j$ and $v_k$, $q_k$ for $1 \le j < k \le M-N+n$. \\
\\
 (3) The following recursive relations between the
intermediate scalar products hold:
\begin{align}
&Q_{M,N,n}(u_1,\dots,u_N|w_1,\dots,w_n|v_1,\dots,v_{M}|h)
|_{w_n=v_{M-N+n}-p-q_{M-N+n}}
=[2p]^{1/2}[2q_{M-N+n}]^{1/2}
\nonumber \\
&\times
\frac{
[h+2(N+n-1)p]^{1/2} [h+2(N+n-1)p+2\overline{q_{M-N+n-1}}]^{1/2}
[h+2(N+n)p+2\overline{q_M}]^{1/2}
}{
[h+2(N+n)p]^{1/2} [h+2(N+n)p+2\overline{q_{M-N+n}}]^{1/2}
[h+2(N+n-1)p+2\overline{q_M}]^{1/2}
}
\nonumber \\
&\times \prod_{j=1}^{M-N+n-1} [v_{M-N+n}-q_{M-N+n}-v_j-q_j]
\prod_{j=M-N+n+1}^{M} [v_{M-N+n}-q_{M-N+n}-v_j+q_j-2p]
\nonumber \\
&\times Q_{M,N,n-1}(u_1,\dots,u_N|w_1,\dots,w_{n-1}|v_1,\dots,v_{M}|h)
. \label{specialpointofintermediatescalarproducts}
\end{align}
\\
 (4) The following evaluation holds for the case $n=0$
\begin{align}
&
Q_{M,N,0}(u_1,\dots,u_N||v_1,\dots,v_{M}|h)
=\frac{[h+2Np+2\overline{q_M}]}{[h+2\overline{q_M}]^{1/2}
[h+2\overline{q_{M-N}}+2Np]^{1/2}} \nonumber \\
&\times \prod_{1 \le j < k \le N} \frac{[u_j-u_k+2p]}{[u_j-u_k]}
\prod_{M-N+1 \le j < k \le M} \frac{[v_k-v_j+q_j+q_k]}{[v_k-v_j+q_j-q_k]}
\prod_{j=1}^N \prod_{k=1}^M [u_j-v_k+p+q_k]
\nonumber \\
&\times
\prod_{j=1}^N [2p]^{1/2} \prod_{j=M-N+1}^M [2q_j]^{1/2}
\mathrm{det}_N
\Bigg(
\frac{[h+(2N-1)p+2\overline{q_M}-q_{M-N+k}+v_{M-N+k}-u_j]}
{[h+2Np+2\overline{q_M}][u_j+p-v_{M-N+k}+q_{M-N+k}]}
\Bigg). \label{intermediateinitialexpression}
\end{align}
\end{proposition}

\begin{figure}[ht]
\includegraphics[width=12cm]{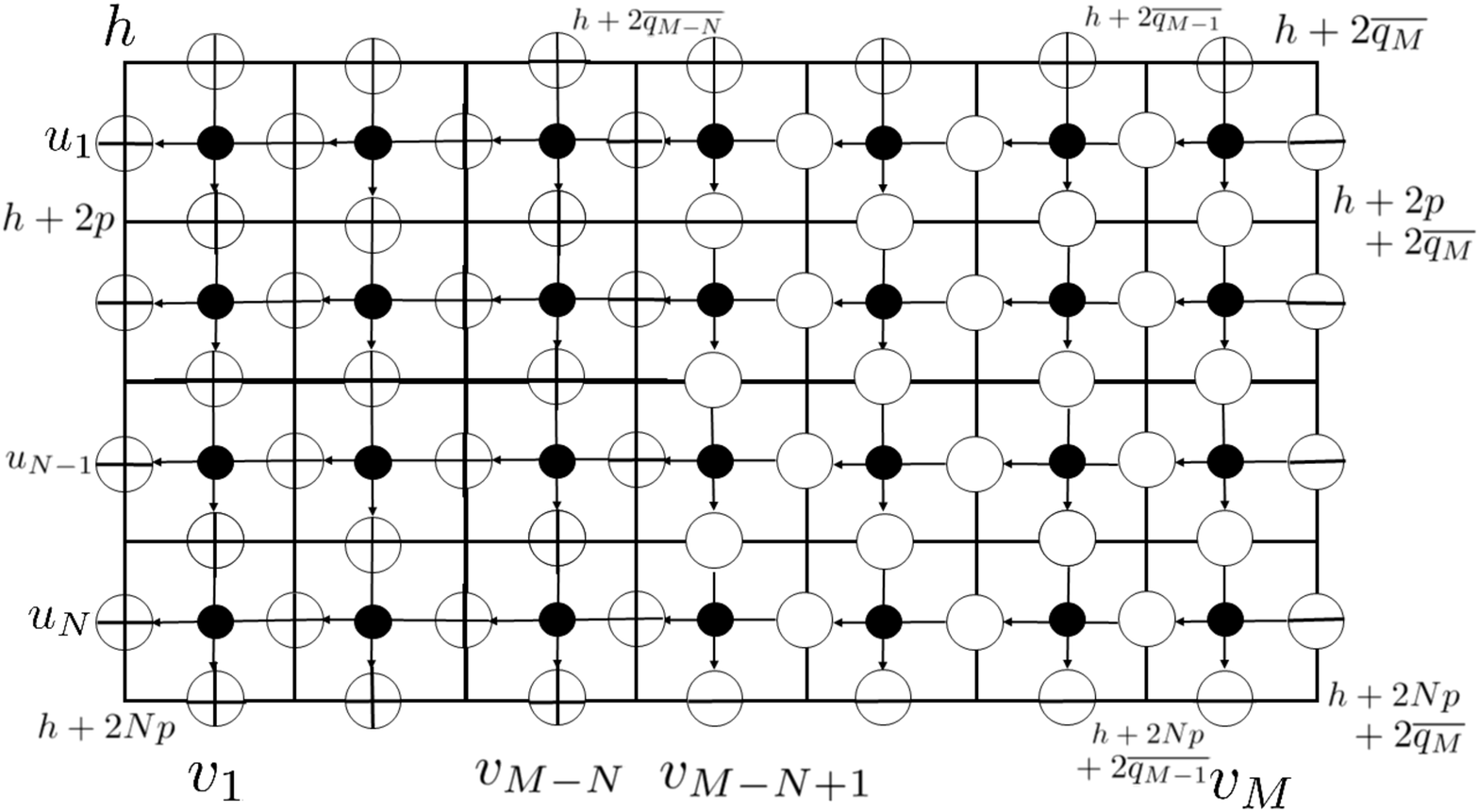}
\caption{The intermediate scalar products
$Q_{M,N,0}(u_1,\dots,u_N||v_1,\dots,v_{M}|h)$.
One can see that the inner states of the left part of the lattice models
are frozen which make contribution
$\prod_{j=1}^N \prod_{k=1}^{M-N} [u_j-v_k+p+q_k]$
to the intermediate scalar products, and the right unfrozen part are
the domain wall boundary partition functions
$Z_N(u_1,\dots,u_N|v_{M-N+1},\dots,v_M|h+2\overline{q_{M-N}})$.
}
\label{picturescalarproductsinitial}
\end{figure}

\begin{proof}
Properties (1), (2) and (3) can be shown
by using standard arguments.

Property (1) can be shown as follows.
We insert the completeness relation into 
the intermediate scalar products
\begin{align}
&Q_{M,N,n}(u_1,\dots,u_N|w_1,\dots,w_n|v_1,\dots,v_{M}|h) \nonumber \\
=&\langle 0^{M-N+n} 1^{N-n}|
C(w_n|v_1,\dots,v_{M}|p|q_1,\dots,q_M|h+2(N+n-1)p) \nonumber \\
&\times \cdots \times
C(w_1|v_1,\dots,v_{M}|p|q_1,\dots,q_M|h+2Np) \nonumber \\
&\times B(u_N|v_1,\dots,v_{M}|p|q_1,\dots,q_M|h+2(N-1)p) \nonumber \\
&\times \cdots \times
B(u_1|v_1,\dots,v_{M}|p|q_1,\dots,q_M|h)|\Omega \rangle_{M} \nonumber \\
=&\sum_{k=1}^{M-N+n}
\langle 0^{M-N+n} 1^{N-n}|
C(w_n|v_1,\dots,v_{M}|p|q_1,\dots,q_M|h+2(N+n-1)p)
|0^{k-1}10^{M-N+n-k}1^{N-n} \rangle
\nonumber \\
&\times
\langle 0^{k-1}10^{M-N+n-k}1^{N-n}|
C(w_{n-1}|v_1,\dots,v_{M}|p|q_1,\dots,q_M|h+2(N+n-2)p)
\times \cdots \times \nonumber \\
&\times C(w_1|v_1,\dots,v_{M}|p|q_1,\dots,q_M|h+2Np)
B(u_N|v_1,\dots,v_{M}|p|q_1,\dots,q_M|h+2(N-1)p)\times \nonumber \\
&\times \cdots \times
B(u_1|v_1,\dots,v_{M}|p|q_1,\dots,q_M|h)|\Omega \rangle_{M}.
\label{decompositionintermediatescalarproducts}
\end{align}
We calculate the matrix elements 
of $C(w_n|v_1,\dots,v_{M}|p|q_1,\dots,q_M|h+2(N+n-1)p)$ based
on its definition to get
\begin{align}
&\langle 0^{M-N+n} 1^{N-n}|
C(w_n|v_1,\dots,v_{M}|p|q_1,\dots,q_M|h+2(N+n-1)p)
|0^{k-1}10^{M-N+n-k}1^{N-n} \rangle \nonumber \\
=&\prod_{j=1}^{k-1}
\frac{[h+2(N+n-1)p+2\overline{q_{j-1}}]^{1/2}
[h+2(N+n)p+2\overline{q_j}]^{1/2}
[w_n-v_j-q_j+p]}{[h+2(N+n)p+2\overline{q_{j-1}}]^{1/2}
[h+2(N+n-1)p+2\overline{q_j}]^{1/2}} \nonumber \\
&\times
\frac{[2p]^{1/2}[2q_k]^{1/2}[w_n-v_k+q_k+p+h+2(N+n-1)p+2\overline{q_{k-1}}]}
{
[h+2(N+n)p+2\overline{q_{k-1}}]^{1/2}[h+2(N+n-1)p+2\overline{q_k}]^{1/2}
}
\prod_{j=k+1}^{M-N+n} [w_n-v_j+p+q_j] \nonumber \\
&\times
\prod_{j=M-N+n+1}^M \frac{
[h+2(N+n-1)p+2\overline{q_{j-1}}]^{1/2}[h+2(N+n)p+2\overline{q_j}]^{1/2}
[w_n-v_j+q_j-p]
}{
[h+2(N+n)p+2\overline{q_{j-1}}]^{1/2}
[h+2(N+n-1)p+2\overline{q_j}]^{1/2}
}. \label{matrixelementsforintermediate}
\end{align}
One can see from
\eqref{matrixelementsforintermediate} that
$Q_{M,N,n}(u_1,\dots,u_N|w_1,\dots,w_n|v_1,\dots,v_{M}|h)$
has
$\displaystyle
\prod_{j=M-N+n+1}^M
[w_n-v_j+q_j-p]
$
as an overall factor
(one can also show this by the graphical representation
of the intermediate scalar products in Figure
\ref{ordinarypictureintermediatescalarproducts},
in which one can show that the
dynamical $R$-matrices of the right part of the bottom row
are already frozen due to the ice-rule of the dynamical $R$-matrix,
and the matrix elements of the $R$-matrices of the frozen parts
contain the factor $\displaystyle
\prod_{j=M-N+n+1}^M
[w_n-v_j+q_j-p]
$).
Let us denote the intermediate scalar products divided by this
factor as $\widetilde{Q}_{M,N,n}(u_1,\dots,u_N|w_1,\dots,w_n|v_1,\dots,v_{M}|h)$:
\begin{align}
&\widetilde{Q}_{M,N,n}(u_1,\dots,u_N|w_1,\dots,w_n|v_1,\dots,v_{M}|h) \nonumber
\\
=&\prod_{j=M-N+n+1}^M
[w_n-v_j+q_j-p]^{-1} Q_{M,N,n}(u_1,\dots,u_N|w_1,\dots,w_n|v_1,\dots,v_{M}|h).
\end{align}
By calculating the quasi-periodicities of
$
\langle 0^{M-N+n} 1^{N-n}|
C(w_n|v_1,\dots,v_{M}|p|q_1,\dots,q_M|h+2(N+n-1)p)
|0^{k-1}10^{M-N+n-k}1^{N-n} \rangle$ with respect to $w_n$, one finds those of
the elliptic functions
$\widetilde{Q}_{M,N,n}(u_1,\dots,u_N|w_1,\dots,w_n|v_1,\dots,v_{M}|h)$
are given by
\begin{align}
&\widetilde{Q}_{M,N,n}(u_1,\dots,u_N|w_1,\dots,w_n+1|v_1,\dots,v_{M}|h)
\nonumber \\
=&(-1)^{M-N+n}
\widetilde{Q}_{M,N,n}(u_1,\dots,u_N|w_1,\dots,w_n|v_1,\dots,v_{M}|h),
\label{intermediateqpfirst} \\
&
\widetilde{Q}_{M,N,n}(u_1,\dots,u_N|w_1,\dots,w_n-i \mathrm{log}({\bf q})/\pi
|v_1,\dots,v_{M}|h)
\nonumber \\
=&(-{\bf q}^{-1})^{M-N+n}
\mathrm{exp} \Bigg(-2 \pi i
\Bigg((M-N+n)w_n+h+(M+N+3n-2)p \nonumber \\
&-\sum_{j=1}^{M-N+n} v_j
+\sum_{j=1}^{M-N+n} q_j
\Bigg)
\Bigg)
\widetilde{Q}_{M,N,n}(u_1,\dots,u_N|w_1,\dots,w_n|v_1,\dots,v_{M}|h),
\label{intermediateqpsecond}
\end{align}
which shows that
$\widetilde{Q}_{M,N,n}(u_1,\dots,u_N|w_1,\dots,w_n|v_1,\dots,v_{M}|h)$
is an elliptic polynomial of $w_n$ in $\Theta_{M-N+n}(\chi)$
of periods $1$ and $\tau=-i \mathrm{log} ({\bf q})/\pi$
with the characters
$\chi(1)=(-1)^{M-N+n}$ and \\
$\displaystyle \chi(\tau)=
\mathrm{exp} \Bigg(-2 \pi i
\Bigg(h+(M+N+3n-2)p
-\sum_{j=1}^{M-N+n} v_j
+\sum_{j=1}^{M-N+n} q_j
\Bigg)
\Bigg)
$. This shows Property (1).

Property (2) can be shown as a consequence of the
the commutation relation between the
vertical monodromy matrices,
which is a standard procedure using the
dynamical Yang-Baxter relation, thus we omit the details.

Property (3) can be shown by substituting
$w_n=v_{M-N+n}-p-q_{M-N+n}$ into
\eqref{decompositionintermediatescalarproducts},
after which only one of the summands $k=M-N+n$
survives. Equivalently, this property can also be shown
by the graphical description of the intermediate
scalar products
(Figure \ref{ordinarypictureintermediatescalarproductsspecialpoint}).

Let us show Property (4).
From its graphical description
(Figure \ref{picturescalarproductsinitial}),
one easily finds that for the case $n=0$,
the intermediate scalar products $Q_{M,N,0}(u_1,\dots,u_N||v_1,\dots,v_{M}|h)$
is just a concatenation of  frozen parts
and the domain wall boundary partition functions
\begin{align}
&Q_{M,N,0}(u_1,\dots,u_N||v_1,\dots,v_{M}|h) \nonumber \\
=&\prod_{j=1}^N \prod_{k=1}^{M-N} [u_j-v_k+p+q_k]
Z_N(u_1,\dots,u_N|v_{M-N+1},\dots,v_M|h+2\overline{q_{M-N}}).
\label{factorizationintermediate}
\end{align}
We insert the factorization formula for the domain wall boundary partition
functions by Foda-Wheeler-Zuparic (\eqref{expressionFWZ}
in Theorem \ref{theoremFWZ})
into the right hand side of \eqref{factorizationintermediate}
to get
\begin{align}
&Q_{M,N,0}(u_1,\dots,u_N||v_1,\dots,v_{M}|h) \nonumber \\
=&\prod_{j=1}^N \prod_{k=1}^{M-N} [u_j-v_k+p+q_k] \nonumber \\
&\times \frac{
\displaystyle \Bigg[h+2\overline{q_{M-N}}+
\sum_{j=M-N+1}^M v_j-\sum_{j=1}^N u_j+Np+\sum_{j=M-N+1}^M q_j \Bigg]}
{[h+2\overline{q_{M}}]^{1/2} [h+2\overline{q_{M-N}}+2Np]^{1/2}} \nonumber \\
&\times
\prod_{j=1}^N [2p]^{1/2}
\prod_{j=M-N+1}^M [2q_j]^{1/2}
\prod_{1 \le j < k \le N}[u_j-u_k+2p]
\prod_{M-N+1 \le j < k \le M}[v_k-v_j+q_j+q_k].
\label{completefactorizationintermediate}
\end{align}
\eqref{completefactorizationintermediate} is already an explicit form
corresponding to the initial condition of the Izergin-Korepin recursion
process between the intermediate scalar products, and is a very compact expression since it is a factorized form. However, we further rewrite it in
a determinant form. Going back to a complicated expression
is because in the next proposition,
we present the explicit determinant form of the intermediate scalar products,
and we have to check that it satisfies the case $n=0$,
which is immediate to see if one rewrites in
the determinant form \eqref{intermediateinitialexpression}.

How to rewrite
\eqref{completefactorizationintermediate} in the determinant form
goes as follows. We set $\lambda=-h-2Np-2\overline{q_M}$,
$z_j=u_j+p$, $w_k=v_{M-N+k}-q_{M-N+k}$
into the Frobenius determinant formula
\begin{align}
\mathrm{det}_N \Bigg(
\frac{[\lambda+z_j-w_k]}{[\lambda][z_j-w_k]}
\Bigg)
=\frac{\displaystyle [\lambda+\sum_{j=1}^N (z_j-w_j)]
\prod_{1 \le j < k \le N}[z_j-z_k][w_k-w_j]
}
{\displaystyle
[\lambda]\prod_{1 \le j , k \le N}[z_j-w_k]
},
\end{align}
to get the following identity
\begin{align}
&\Bigg[ h+2\overline{q_{M-N}}+
\sum_{j=M-N+1}^M v_j-\sum_{j=1}^N u_j+Np+\sum_{j=M-N+1}^M q_j \Bigg]
\nonumber \\
=&
\frac{\displaystyle [h+2Np+2\overline{q_M}]
\prod_{j=1}^N \prod_{k=M-N+1}^M [u_j+p-v_k+q_k]}
{\displaystyle \prod_{1 \le j < k \le N} [u_j-u_k]
\prod_{M-N+1 \le j < k \le M} [v_k-v_j-q_k+q_j]
} \nonumber \\
&\times \mathrm{det}_N
\Bigg(
\frac{[h+(2N-1)p+2\overline{q_M}-q_{M-N+k}+v_{M-N+k}-u_j]}
{[h+2Np+2\overline{q_M}][u_j+p-v_{M-N+k}+q_{M-N+k}]}
\Bigg). \label{tobeuse}
\end{align}
Substituting \eqref{tobeuse} into the right hand side of
\eqref{completefactorizationintermediate}, we get
\begin{align}
&Q_{M,N,0}(u_1,\dots,u_N||v_1,\dots,v_{M}|h) \nonumber \\
=&\frac{1}
{[h+2\overline{q_{M}}]^{1/2} [h+2\overline{q_{M-N}}+2Np]^{1/2}} 
\prod_{j=1}^N \prod_{k=1}^{M-N} [u_j-v_k+p+q_k] \nonumber \\
&\times 
\frac{\displaystyle [h+2Np+2\overline{q_M}]
\prod_{j=1}^N \prod_{k=M-N+1}^M [u_j+p-v_k+q_k]}
{\displaystyle \prod_{1 \le j < k \le N} [u_j-u_k]
\prod_{M-N+1 \le j < k \le M} [v_k-v_j-q_k+q_j]
} \nonumber \\
&\times \mathrm{det}_N
\Bigg(
\frac{[h+(2N-1)p+2\overline{q_M}-q_{M-N+k}+v_{M-N+k}-u_j]}
{[h+2Np+2\overline{q_M}][u_j+p-v_{M-N+k}+q_{M-N+k}]}
\Bigg)
\nonumber \\
&\times
\prod_{j=1}^N [2p]^{1/2}
\prod_{j=M-N+1}^M [2q_j]^{1/2}
\prod_{1 \le j < k \le N}[u_j-u_k+2p]
\prod_{M-N+1 \le j < k \le M}[v_k-v_j+q_j+q_k] \nonumber \\
=&\frac{[h+2Np+2\overline{q_M}]}{[h+2\overline{q_M}]^{1/2}
[h+2\overline{q_{M-N}}+2Np]^{1/2}} \nonumber \\
&\times \prod_{1 \le j < k \le N} \frac{[u_j-u_k+2p]}{[u_j-u_k]}
\prod_{M-N+1 \le j < k \le M} \frac{[v_k-v_j+q_j+q_k]}{[v_k-v_j+q_j-q_k]}
\prod_{j=1}^N \prod_{k=1}^M [u_j-v_k+p+q_k]
\nonumber \\
&\times
\prod_{j=1}^N [2p]^{1/2} \prod_{j=M-N+1}^M [2q_j]^{1/2}
\mathrm{det}_N
\Bigg(
\frac{[h+(2N-1)p+2\overline{q_M}-q_{M-N+k}+v_{M-N+k}-u_j]}
{[h+2Np+2\overline{q_M}][u_j+p-v_{M-N+k}+q_{M-N+k}]}
\Bigg),
\end{align}
and Property (4) is proved.

\end{proof}

The next thing to do is to find the explicit forms of the intermediate
scalar products satisfying all the properties in
Proposition \ref{aboutgeneralizedscalarproducts}.
One can show the following determinant representation.
\begin{theorem}
\label{theoremintermediatescalarproducts}
The intermediate scalar products
$Q_{M,N,n}(u_1,\dots,u_N|w_1,\dots,w_n|v_1,\dots,v_{M}|h)$
have the following determinant form:
\begin{align}
&
Q_{M,N,n}(u_1,\dots,u_N|w_1,\dots,w_n|v_1,\dots,v_{M}|h) \nonumber \\
=&D_{M,N,n}
\prod_{1 \le j < k \le N} \frac{[u_j-u_k+2p]}{[u_j-u_k]}
\prod_{M-N+n+1 \le j < k \le M} \frac{[v_k-v_j+q_j+q_k]}{[v_k-v_j+q_j-q_k]}
\nonumber \\
&\times
\prod_{1 \le j < k \le n} \frac{[w_j-w_k-2p]}{[w_j-w_k]}
\mathrm{det}_N
\Bigg(
P_{M,N,n}(u_1,\dots,u_N|w_1,\dots,w_n|v_1,\dots,v_{M}|h)
\Bigg), \label{intermediatedeterminant}
\end{align}
where $D_{M,N,n}$ is given by
\begin{align}
D_{M,N,n}=&[2p]^{(N+n)/2} \prod_{j=n+1}^N [2q_{M-N+j}]^{1/2} \nonumber \\
&\times
\frac{
[h+2Np]^{1/2} [h+2Np+2\overline{q_{M}}]^{1/2}
[h+2(N+n)p+2\overline{q_M}]^{1/2}
}{[h+2\overline{q_M}]^{1/2}[h+2(N+n)p]^{1/2}
[h+2(N+n)p+2\overline{q_{M-N+n}}]^{1/2}
},
\end{align}
and $P_{M,N,n}(u_1,\dots,u_N|w_1,\dots,w_n|v_1,\dots,v_{M}|h)$
is an $N \times N$ matrix whose matrix elements are given by
\begin{align}
&P_{M,N,n}(u_1,\dots,u_N|w_1,\dots,w_n|v_1,\dots,v_{M}|h)
_{jk} \nonumber \\
=&
\begin{cases}
\displaystyle \prod_{\ell=M-N+n+1}^M \frac{\displaystyle [w_j-v_\ell-p+q_\ell]}
{\displaystyle [w_j-v_\ell+p+q_\ell]}
\frac{\frac{[w_j+h+2Np-u_k]}{[h+2Np]}a(w_j)d(u_k)
-\frac{[w_j+h+2Np-u_k+2\overline{q_M}]}{[h+2Np+2\overline{q_M}]}a(u_k)d
(w_j)
}{[u_k-w_j]},
\\
\text{ ($1\le j \le n$)}  \\
\displaystyle
\frac{[h+(2N-1)p+2\overline{q_M}-q_{M-N+j}+v_{M-N+j}-u_k]}{[h+2Np+2\overline{q_M}]}
\prod_{\substack{\ell=1 \\ \ell \neq M-N+j}}^M
[u_k-v_\ell+p+q_\ell],
\\
\text{ ($n+1\le j \le N$)}
\end{cases}.
\label{intermediatescalarproductsmatrixelements}
\end{align}
\end{theorem}

\begin{proof}
One can check directly that
the right hand side of \eqref{intermediatedeterminant}
satisfies Properties (1), (2), (3) and (4)
in Proposition \ref{aboutgeneralizedscalarproducts}.
We give some comments.
Let us denote the right hand side of
\eqref{intermediatedeterminant} as $R_{M,N,n}(u_1,\dots,u_N|w_1,\dots,w_n|v_1,\dots,v_{M}|h)$ and set
\begin{align}
&\widetilde{R}_{M,N,n}(u_1,\dots,u_N|w_1,\dots,w_n|v_1,\dots,v_{M}|h)
\nonumber \\
=&\displaystyle \prod_{j=M-N+n+1}^M [w_n-v_j+q_j-p]^{-1}
R_{M,N,n}(u_1,\dots,u_N|w_1,\dots,w_n|v_1,\dots,v_{M}|h).
\end{align}
Property (1) can be shown by
calculating
the quasiperiodicites of the function \\
$\widetilde{R}_{M,N,n}(u_1,\dots,u_N|w_1,\dots,w_n|v_1,\dots,v_{M}|h)$
 with respect to $w_n$.
Expanding the determinant
in the right hand side of \eqref{intermediatedeterminant},
recalling that $a(u)$ and $d(u)$ are defined as
\begin{align}
a(u)=\prod_{\ell=1}^M [u-v_\ell+p+q_\ell], \ \ \
d(u)=\prod_{\ell=1}^M [u-v_\ell+p-q_\ell],
\end{align}
concentrating on
the factors depending on $w_n$,
one finds
\begin{align}
&\widetilde{R}_{M,N,n}(u_1,\dots,u_N|w_1,\dots,w_n+1|v_1,\dots,v_{M}|h)
\nonumber \\
=&(-1)^{M-N+n}
\widetilde{R}_{M,N,n}(u_1,\dots,u_N|w_1,\dots,w_n|v_1,\dots,v_{M}|h), \\
&
\widetilde{R}_{M,N,n}(u_1,\dots,u_N|w_1,\dots,w_n-
i \mathrm{log}({\bf q})/\pi|v_1,\dots,v_{M}|h)
\nonumber \\
=&(-{\bf q}^{-1})^{M-N+n}
\mathrm{exp} \Bigg(-2 \pi i
\Bigg((M-N+n)w_n+h+(M+N+3n-2)p \nonumber \\
&-\sum_{j=1}^{M-N+n} v_j
+\sum_{j=1}^{M-N+n} q_j
\Bigg)
\Bigg)
\widetilde{R}_{M,N,n}(u_1,\dots,u_N|w_1,\dots,w_n|v_1,\dots,v_{M}|h),
\end{align}
which are exactly the same with those for
$\widetilde{Q}_{M,N,n}(u_1,\dots,u_N|w_1,\dots,w_n+1|v_1,\dots,v_{M}|h)$
\eqref{intermediateqpfirst} and
\eqref{intermediateqpsecond}.

One can also show that
$\widetilde{R}_{M,N,n}(u_1,\dots,u_N|w_1,\dots,w_n+1|v_1,\dots,v_{M}|h)$
has apparent singularities at $w_n=u_j$, $j=1,\dots,n$
coming from the zeroes of the denominators of the matrix elements
\eqref{intermediatescalarproductsmatrixelements}, which cancel
with the corresponding zeroes of the numerators.
There are also apparent singularities at $w_n=w_j$, $j=1,\dots,n-1$
and $w_n=v_\ell-p-q_\ell$, $\ell=M-N+n+1,\dots,M$.
Again, one finds that in these cases two rows
of the matrix $P_{M,N,n}(u_1,\dots,u_N|w_1,\dots,w_n|v_1,\dots,v_{M}|h)$
become proportional when taking the limits
$w_n \to w_j$, $j=1,\dots,n-1$ or
$w_n \to v_\ell-p-q_\ell$, $\ell=M-N+n+1,\dots,M$,
hence there are no singularities and
$\widetilde{R}_{M,N,n}(u_1,\dots,u_N|w_1,\dots,w_n+1|v_1,\dots,v_{M}|h)$
is an elliptic polynomial as a function of $w_n$.

Property (2) can be easily checked
by recalling that $\overline{q_M}$ and $\overline{q_{M-N+n}}$ which appear
in
$R_{M,N,n}(u_1,\dots,u_N|w_1,\dots,w_n|v_1,\dots,v_{M}|h)$
are defined as
$\displaystyle \overline{q_M}=\sum_{j=1}^M q_j$ and
$\displaystyle \overline{q_{M-N+n}}=\sum_{j=1}^{M-N+n} q_j$
from which one can see that they are
invariant under the exchange of $q_j$ and $q_k$ for $1 \le j < k \le M-N+n$.
Property (3) can be checked by a long and
tedious but straightforward computation.
We remark that expanding the determinant in the right hand side
of \eqref{intermediatedeterminant}
$\displaystyle \mathrm{det}_N
\Bigg(
P_{M,N,n}(u_1,\dots,u_N|w_1,\dots,w_n|v_1,\dots,v_{M}|h)
\Bigg)$ based on its definition, and
rewriting the prefactor $D_{M,N,n}$
in the function $R_{M,N,n}(u_1,\dots,u_N|w_1,\dots,w_n|v_1,\dots,v_{M}|h)$ as
\begin{align}
&D_{M,N,n} \nonumber \\
=&
\frac{[h+2Np+2\overline{q_{M}}]
}{[h+2\overline{q_M}]^{1/2}[h++2\overline{q_{M-N}}+2Np]^{1/2}
}
\prod_{j=1}^N [2p]^{1/2} [2q_{M-N+j}]^{1/2}
\prod_{j=1}^n \frac{[2p]^{1/2}}{[2q_{M-N+j}]^{1/2}}
\nonumber \\
\times&
\prod_{j=1}^n
\frac{
[h+2(N+j-1)p]^{1/2} [h+2(N+j-1)p+2\overline{q_{M-N+j-1}}]^{1/2}
[h+2(N+j)p+2\overline{q_M}]^{1/2}
}{[h+2(N+j)p]^{1/2}[h+2(N+j-1)p+2\overline{q_M}]^{1/2}
[h+2(N+j)p+2\overline{q_{M-N+j}}]^{1/2}
},
\end{align}
makes things easier to check Property (3).

Property (4) can be checked immediately
by setting $n=0$ in the right hand side of
\eqref{intermediatedeterminant}.
\end{proof}

As a consequence of Theorem \ref{theoremintermediatescalarproducts},
one gets the determinant formula for the scalar products \eqref{determinantscalarproducts} by specializing \eqref{intermediatedeterminant} to $n=N$,
hence we have proved Theorem \ref{maintheoremscalarproducts}.

\section{Elliptic Cauchy formula}\label{sec4}

We derive the Cauchy formula for elliptic symmetric functions
by combining the determinant formula for the scalar products
proved in the last section
with another evaluation based on the correspondence
between the wavefunctions and the symmetric functions.
Let us first recall the correspondence \cite{Motr}.
A detailed proof of the correspondence can also be found for
the closely related Okado-Deguchi-Fujii-Martin
\cite{Okado,DF,DM} (elliptic Perk-Schultz) model \cite{Moelliptic}.

We introduce a class of partition functions
$W_{M,N}(u_1,\dots,u_N|v_1,\dots,v_M|x_1,\dots,x_N|h)$
defined as the matrix elements of the product of the $B$-operators
\eqref{Boperator} as follows:
\begin{align}
&W_{M,N}(u_1,\dots,u_N|v_1,\dots,v_M|x_1,\dots,x_N|h)
\nonumber \\
=&\langle x_1 \cdots x_N|
B(u_N|v_1,\dots,v_M|p|q_1,\dots,q_M|h+2(N-1)p) \nonumber \\
&\times \cdots \times
B(u_2|v_1,\dots,v_M|p|q_1,\dots,q_M|h+2p)
B(u_1|v_1,\dots,v_M|p|q_1,\dots,q_M|h)|\Omega \rangle_M,
\label{wavefunction}
\end{align}
where $\langle x_1 \cdots x_N|$ are the dual $N$-particle states
\begin{align}
\langle x_1 \cdots x_N|
&=(_1 \langle 0| \otimes \cdots \otimes {}_M \langle 0|)
\prod_{j=1}^N \sigma^+_{x_j}
\in \mathcal{F}_1^* \otimes \cdots \otimes \mathcal{F}_M^*
, \label{ordinarydualparticleconfiguration}
\end{align}
which are states labelling the configurations
of particles
$1 \le x_1 < x_2 < \cdots < x_N \le M$.
We call this class of partition functions as wavefunctions in this paper
since it is an analogue of wavefunctions of integrable vertex models.

We also define another class of wavefunctions $V_{M,N}(u_1,\dots,u_N|v_1,\dots,v_M|x_1,\dots,x_N|h)$
as matrix elements of the $C$-operators \eqref{Coperator} as
\begin{align}
&V_{M,N}(u_1,\dots,u_N|v_1,\dots,v_M|x_1,\dots,x_N|h)
\nonumber \\
=&{}_M \langle \Omega|
C(u_N|v_1,\dots,v_M|p|q_1,\dots,q_M|h+2(N-1)p) \nonumber \\
&\times \cdots \times
C(u_2|v_1,\dots,v_M|p|q_1,\dots,q_M|h+2p)
C(u_1|v_1,\dots,v_M|p|q_1,\dots,q_M|h)|x_1 \cdots x_N \rangle,
\label{dualwavefunction}
\end{align}
where $|x_1 \cdots x_N \rangle$
are the $N$-particle states
\begin{align}
|x_1 \cdots x_N \rangle
&=
\prod_{j=1}^N \sigma^-_{x_j}
(|0 \rangle_1 \otimes \cdots \otimes {}_M |0 \rangle_M)
\in \mathcal{F}_1 \otimes \cdots \otimes \mathcal{F}_M
.
\end{align}

\begin{figure}[ht]
\includegraphics[width=12cm]{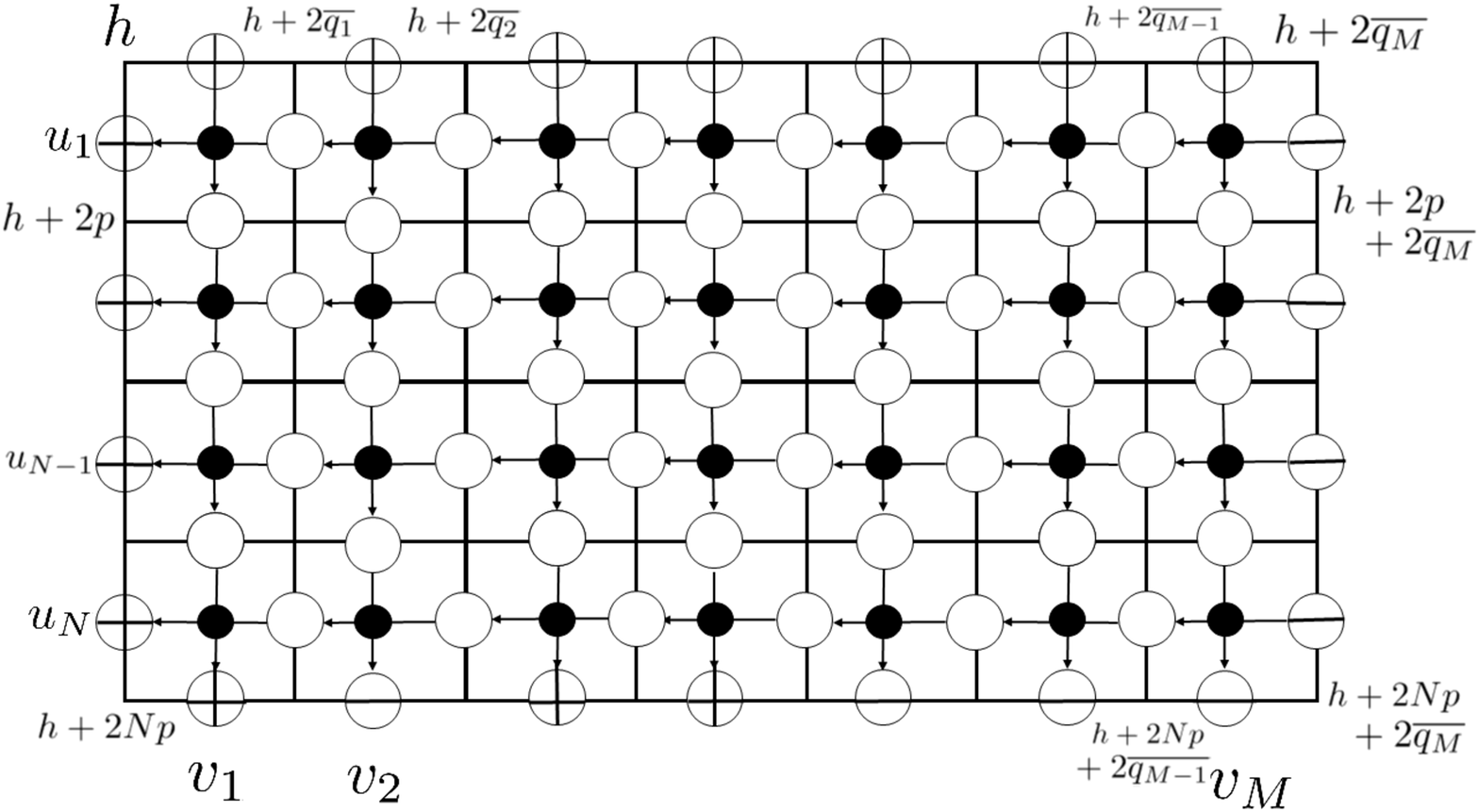}
\caption{The wavefunctions
$W_{M,N}(u_1,\dots,u_N|v_1,\dots,v_M|x_1,\dots,x_N|h)$
\protect\eqref{wavefunction}.
The figure illustrates the case $M=7$, $N=4$,
$x_1=2$, $x_2=5$, $x_3=6$, $x_4=7$.
}
\label{picturewavefunctions}
\end{figure}

\begin{figure}[ht]
\includegraphics[width=12cm]{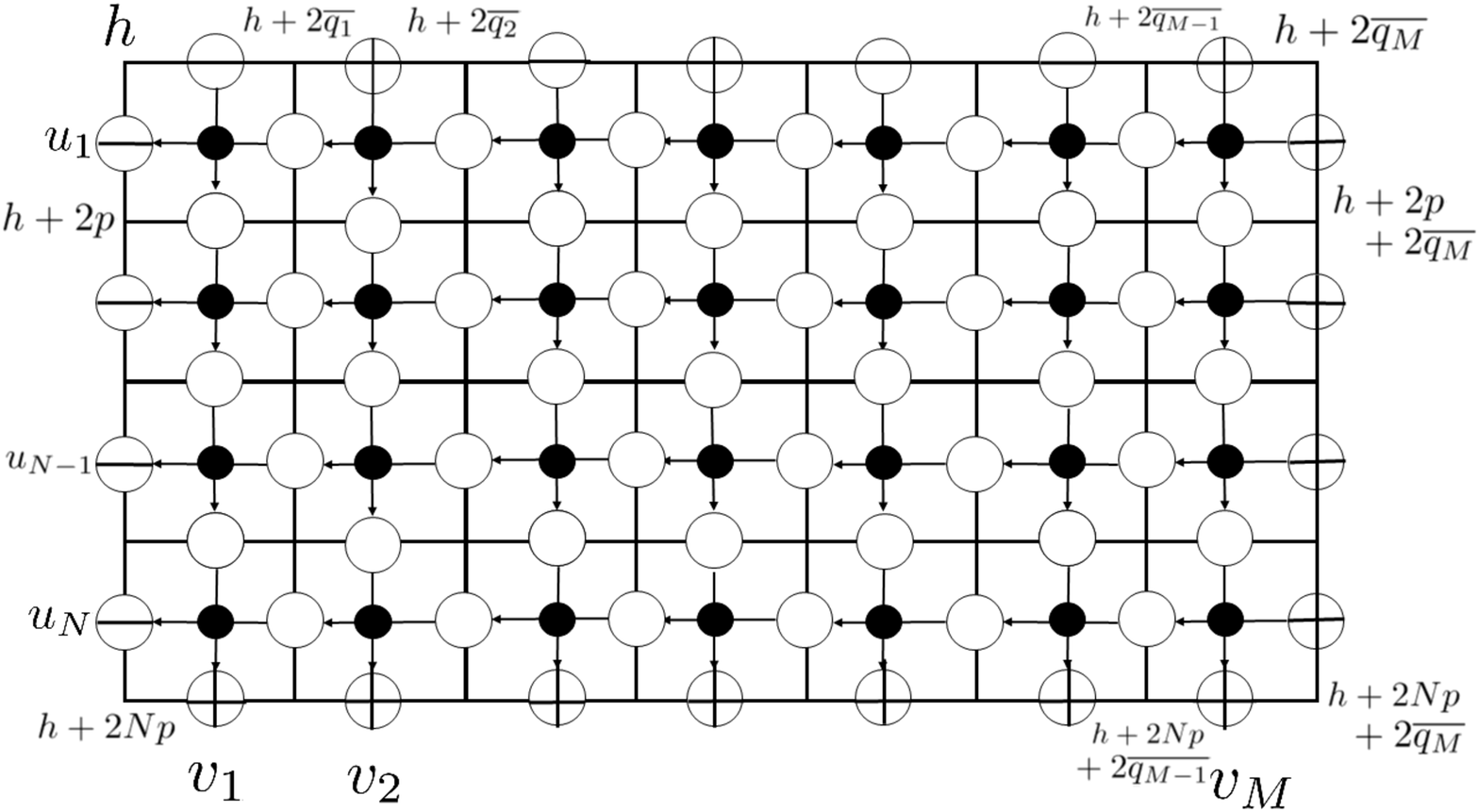}
\caption{The wavefunctions
$V_{M,N}(u_1,\dots,u_N|v_1,\dots,v_M|x_1,\dots,x_N|h)$
\protect\eqref{dualwavefunction}.
The figure illustrates the case $M=7$, $N=4$,
$x_1=1$, $x_2=3$, $x_3=5$, $x_4=6$.
}
\label{picturewavefunctionstwo}
\end{figure}

See Figures \ref{picturewavefunctions} and 
\ref{picturewavefunctionstwo}
for graphical representations
of the wavefunctions \eqref{wavefunction}, \eqref{dualwavefunction}.

The wavefunctions \eqref{wavefunction}, \eqref{dualwavefunction}
can be explicitly
expressed using deformed elliptic Vandermonde determinants
and elliptic symmetric functions defined below.

\begin{definition}
We define the following elliptic Schur function \\
$S_{M,N}(u_1,\dots,u_N|v_1,\dots,v_M|x_1,\dots,x_N|h)$
which depends on the symmetric variables \\
$u_1,\dots,u_N$,
two sets of complex parameters $v_1,\dots,v_M$ and
$q_1,\dots,q_M$, two complex parameters $h$, $p$
and integers $x_1,\dots,x_N$ satisfying
$1 \le x_1 < \cdots < x_N \le M$,
\begin{align}
&S_{M,N}(u_1,\dots,u_N|v_1,\dots,v_M|x_1,\dots,x_N|h)
\nonumber \\
=&\sum_{\sigma \in S_N}
\prod_{1 \le j < k \le N} \frac{1}{[u_{\sigma(j)}-u_{\sigma(k)}]}
\prod_{j=1}^N \prod_{k=x_j+1}^M [u_{\sigma(j)}-v_k-q_k+p]
\nonumber \\
&\times
\prod_{j=1}^N
\frac{[h+2jp+2\overline{q_M}]^{1/2} [2p]^{1/2} [2q_{x_j}]^{1/2}
}
{[h+2(j-1)p+2\overline{q_M}]^{1/2}[h+2Np+2\overline{q_{x_j-1}}]^{1/2}
[h+2Np+2\overline{q_{x_j}}]^{1/2}}
\nonumber \\
&\times
\prod_{j=1}^N [-u_{\sigma(j)}+v_{x_j}+h+(2N-1)p+q_{x_j}+2\overline{q_{x_j-1}}]
\prod_{j=1}^N \prod_{k=1}^{x_j-1}
[u_{\sigma(j)}-v_k+p+q_k],
\label{ordinaryrighthandside} \\
=&\prod_{1 \le j < k \le N} \frac{1}{[u_{j}-u_{k}]}
\mathrm{det}_N (f_{x_j}(u_k|v_1,\dots,v_M)), \end{align}
\begin{align}
&f_{x_j}(u|v_1,\dots,v_M) \nonumber \\
=&
\frac{[h+2jp+2\overline{q_M}]^{1/2} [2p]^{1/2} [2q_{x_j}]^{1/2}
}
{[h+2(j-1)p+2\overline{q_M}]^{1/2}[h+2Np+2\overline{q_{x_j-1}}]^{1/2}
[h+2Np+2\overline{q_{x_j}}]^{1/2}} \nonumber \\
&\times [-u+v_{x_j}+h+(2N-1)p+q_{x_j}+2\overline{q_{x_j-1}}]
\prod_{k=1}^{x_j-1} [u-v_k+p+q_k]
\prod_{k=x_j+1}^M [u-v_k+p-q_k].
\end{align}
Recall that $\overline{q_j}$ is defined as
$\overline{q_j}=\sum_{k=1}^j q_k$.

We also define another elliptic Schur function
$T_{M,N}(u_1,\dots,u_N|v_1,\dots,v_M|x_1,\dots,x_N|h)$
which depends on the symmetric variables
$u_1,\dots,u_N$,
two sets of complex parameters $v_1,\dots,v_M$ and
$q_1,\dots,q_M$, two complex parameters $h$, $p$
and integers $x_1,\dots,x_N$ satisfying
$1 \le x_1 < \cdots < x_N \le M$,
\begin{align}
&T_{M,N}(u_1,\dots,u_N|v_1,\dots,v_M|x_1,\dots,x_N|h)
\nonumber \\
=&\sum_{\sigma \in S_N}
\prod_{1 \le j < k \le N} \frac{1}{[u_{\sigma(j)}-u_{\sigma(k)}]}
\prod_{j=1}^N \prod_{k=x_j+1}^M [u_{\sigma(j)}-v_k+q_k+p]
\nonumber \\
&\times
\prod_{j=1}^N
\frac{[h+2(j-1)p]^{1/2} [2p]^{1/2} [2q_{x_j}]^{1/2}
}
{[h+2jp]^{1/2}[h+2\overline{q_{x_j-1}}]^{1/2}
[h+2\overline{q_{x_j}}]^{1/2}}
\nonumber \\
&\times
\prod_{j=1}^N [u_{\sigma(j)}-v_{x_j}+h+p+q_{x_j}
+2\overline{q_{x_j-1}}]
\prod_{j=1}^N \prod_{k=1}^{x_j-1}
[u_{\sigma(j)}-v_k+p-q_k],
\label{dualordinaryrighthandside} \\
=&\prod_{1 \le j < k \le N} \frac{1}{[u_{j}-u_{k}]}
\mathrm{det}_N (h_{x_j}(u_k|v_1,\dots,v_M)), \end{align}
\begin{align}
&h_{x_j}(u|v_1,\dots,v_M) \nonumber \\
=&
\frac{[h+2(j-1)p]^{1/2} [2p]^{1/2} [2q_{x_j}]^{1/2}
}
{[h+2jp]^{1/2}[h+2\overline{q_{x_j-1}}]^{1/2}
[h+2\overline{q_{x_j}}]^{1/2}} \nonumber \\
&\times
[u-v_{x_j}+h+p+q_{x_j}+2\overline{q_{x_j-1}}]
\prod_{k=1}^{x_j-1} [u-v_k+p-q_k]
\prod_{k=x_j+1}^M [u-v_k+p+q_k].
\end{align}

\end{definition}

The wavefunctions of the
elliptic Felderhof model can be expressed as
products of one-parameter deformations of the
elliptic Vandermonde determinant
and the elliptic Schur functions
$S_{M,N}(u_1,\dots,u_N|v_1,\dots,v_M|x_1,\dots,x_N|h)$
and
$T_{M,N}(u_1,\dots,u_N|v_1,\dots,v_M|x_1,\dots,x_N|h)$ defined above.
We present the correspondence below.

\begin{theorem} \label{maintheoremstatement}
The wavefunctions of the elliptic Felderhof model
\\
$W_{M,N}(u_1,\dots,u_N|v_1,\dots,v_M|x_1,\dots,x_N|h)$
is explicitly expressed as the
product of a one-parameter deformation of an
elliptic Vandermonde determinant
$\displaystyle \prod_{1 \le j < k \le N} [u_j-u_k+2p]$ and 
the elliptic Schur functions
$S_{M,N}(u_1,\dots,u_N|v_1,\dots,v_M|x_1,\dots,x_N|h)$
\begin{align}
&W_{M,N}(u_1,\dots,u_N|v_1,\dots,v_M|x_1,\dots,x_N|h) \nonumber \\
=&
\prod_{1 \le j < k \le N} [u_j-u_k+2p]
S_{M,N}(u_1,\dots,u_N|v_1,\dots,v_M|x_1,\dots,x_N|h).
\label{maintheorem}
\end{align}

The wavefunctions
$V_{M,N}(u_1,\dots,u_N|v_1,\dots,v_M|x_1,\dots,x_N|h)$
is explicitly expressed as the
product of a one-parameter deformation of an
elliptic Vandermonde determinant
$\displaystyle \prod_{1 \le j < k \le N} [u_j-u_k-2p]$ and 
the elliptic Schur functions
$T_{M,N}(u_1,\dots,u_N|v_1,\dots,v_M|x_1,\dots,x_N|h)$
\begin{align}
&V_{M,N}(u_1,\dots,u_N|v_1,\dots,v_M|x_1,\dots,x_N|h) \nonumber \\
=&
\prod_{1 \le j < k \le N} [u_j-u_k-2p]
T_{M,N}(u_1,\dots,u_N|v_1,\dots,v_M|x_1,\dots,x_N|h).
\label{dualmaintheorem}
\end{align}

\end{theorem}

The relation \eqref{maintheorem} is shown in \cite{Motr}.
In the Appendix, we give a proof of \eqref{dualmaintheorem}
in some detail.

Now, combining the determinant formula for the
scalar products (Theorem \ref{maintheoremscalarproducts})
and the correspondence between the wavefunctions
and the elliptic Schur functions (Theorem \ref{maintheoremstatement}),
one can derive the Cauchy formula
for the elliptic Schur functions.

\begin{theorem} \label{inhomogeneouspairing}
We have the following Cauchy formula for the
elliptic Schur functions
\begin{align}
\sum_{1 \le x_1 < x_2 < \cdots < x_N \le M}
&S_{M,N}(u_1,\dots,u_N|v_1,\dots,v_{M}|x_1,\dots,x_N|h) \nonumber \\
\times&T_{M,N}(w_1,\dots,w_N|v_1,\dots,v_{M}|x_1,\dots,x_N|h+2Np) \nonumber \\
=&[2p]^N \frac{[h+2Np]^{1/2}[h+2Np+2\overline{q_M}]^{1/2}}
{[h+4Np]^{1/2} [h+2\overline{q_M}]^{1/2}}
\frac{1}{\prod_{1 \le j < k \le N}[u_j-u_k][w_j-w_k]} \nonumber \\
&\times \mathrm{det}_N
\Bigg(
\frac{\frac{[w_j+h+2Np-u_k]}{[h+2Np]}a(w_j)d(u_k)
-\frac{[w_j+h+2Np-u_k+2\overline{q_M}]}{[h+2Np+2\overline{q_M}]}a(u_k)d(w_j)
}{[u_k-w_j]}
\Bigg).
\label{pairing}
\end{align}
\end{theorem}

\begin{figure}[ht]
\includegraphics[width=15cm]{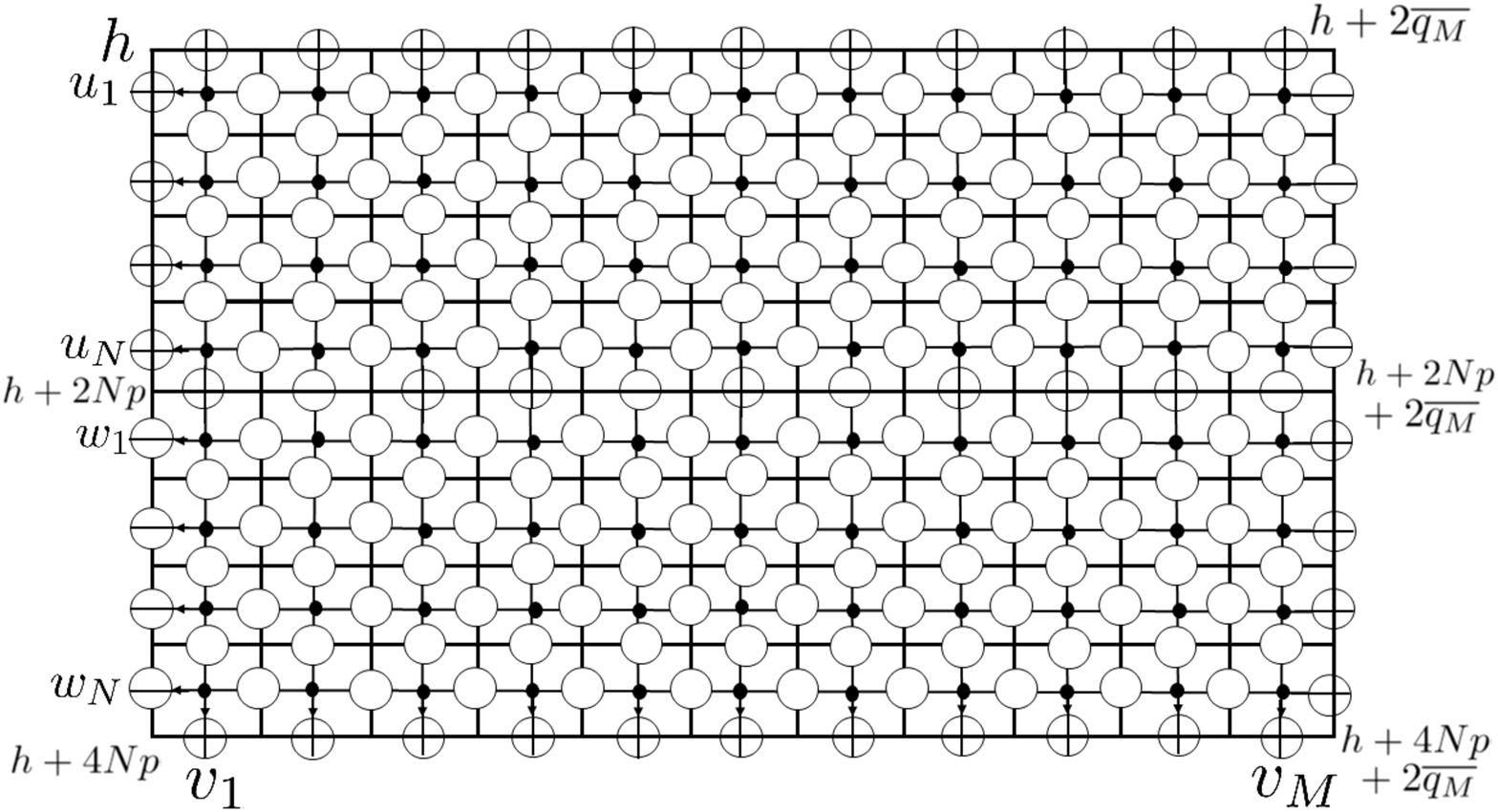}
\caption{A graphical representation of a summand
corresponding to $x_1=2$, $x_2=5$, $x_3=7$, $x_4=11$
in the decomposition of the scalar products
into the sum over products of the wavefunctions
\protect\eqref{forpairing}.
}
\label{ordinarypicturedecomposition}
\end{figure}

\begin{proof}
The scalar products \eqref{scalarproducts}
can be evaluated by inserting
the completeness relation
\begin{align}
\sum_{1 \le x_1 < x_2 < \cdots < x_N \le M}
|x_1 \cdots x_N \rangle \langle x_1 \cdots x_N |=\mathrm{Id},
\end{align}
and using the correspondence between
the wavefunctions and the elliptic Schur functions
\eqref{maintheorem}, \eqref{dualmaintheorem} in
Theorem \ref{maintheoremstatement} as
\begin{align}
&Q_{M,N}(u_1,\dots,u_N|w_1,\dots,w_N|v_1,\dots,v_{M}|h)
\nonumber \\
=&{}_{M} \langle \Omega|
C(w_N|v_1,\dots,v_{M}|p|q_1,\dots,q_M|h+2(2N-1)p) \nonumber \\
&\times \cdots \times
C(w_1|v_1,\dots,v_{M}|p|q_1,\dots,q_M|h+2Np) \nonumber \\
&\times B(u_N|v_1,\dots,v_{M}|p|q_1,\dots,q_M|h+2(N-1)p) \nonumber \\
&\times \cdots \times
B(u_1|v_1,\dots,v_{M}|p|q_1,\dots,q_M|h)|\Omega \rangle_{M}, \nonumber \\
=&\sum_{1 \le x_1 < x_2 < \cdots < x_N \le M} {}_{M} \langle \Omega|
C(w_N|v_1,\dots,v_{M}|p|q_1,\dots,q_M|h+2(2N-1)p) \nonumber \\
&\times \cdots \times
C(w_1|v_1,\dots,v_{M}|p|q_1,\dots,q_M|h+2Np)
|x_1 \cdots x_N \rangle
\nonumber \\
&\times \langle x_1 \cdots x_N | B(u_N|v_1,\dots,v_{M}|p|q_1,\dots,q_M|h+2(N-1)p) \nonumber \\
&\times \cdots \times
B(u_1|v_1,\dots,v_{M}|p|q_1,\dots,q_M|h)|\Omega \rangle_{M}, \nonumber \\
=&\sum_{1 \le x_1 < x_2 < \cdots < x_N \le M}
V_{M,N}(w_1,\dots,w_N|v_1,\dots,v_{M}|x_1,\dots,x_N|h+2Np)
\nonumber \\
&\times
W_{M,N}(u_1,\dots,u_N|v_1,\dots,v_{M}|x_1,\dots,x_N|h)
\nonumber \\
=&\sum_{1 \le x_1 < x_2 < \cdots < x_N \le M}
\prod_{1 \le j < k \le N} [w_j-w_k-2p]
T_{M,N}(w_1,\dots,w_N|v_1,\dots,v_{M}|x_1,\dots,x_m|h+2Np) \nonumber \\
&\times
\prod_{1 \le j < k \le N} [u_j-u_k+2p]
S_{M,N}(u_1,\dots,u_N|v_1,\dots,v_{M}|x_1,\dots,x_N|h). \label{forpairing}
\end{align}
See Figure \ref{ordinarypicturedecomposition} for a graphical description
of the decomposition \eqref{forpairing}.
The Cauchy formula for the elliptic Schur functions
\eqref{pairing} follows from comparing \eqref{forpairing}
with the direct evaluation which gives the
determinant formula for the scalar products
\eqref{determinantscalarproducts} (Theorem \ref{maintheoremscalarproducts}).

\end{proof}

\section{Conclusion}
In this paper, we examined the scalar products
of the elliptic Felderhof model introduced
by Foda-Wheeler-Zuparic \cite{FWZ}, which is an elliptic
extension of the face-type Felderhof model \cite{Felderhof} of Deguchi-Akutsu \cite{DA}.
By applying the
Izergin-Korepin technique developed by Wheeler \cite{Wheeler} to analyze
the scalar produts, we derived the determinant formula
for the scalar products of the elliptic Felderhof model
by constructing the explicit determinant forms of the intermediate
scalar products, which connects the scalar products
and the domain wall boundary partition functions.

The scalar products can be evaluated in another way
as a sum over products of the wavefunctions.
For the case of the elliptic Felderhof models,
the wavefunctions are expressed as
a deformed elliptic Vandermonde determinant
and elliptic Schur functions, which can be shown for example by
the recently-developed Izergin-Korepin technique
to analyze the wavefunctions \cite{MoIK,Moelliptic,Motr}.
By combining this way of evaluation with the direct evaluation
which gives the determinant formula,
we obtained the Cauchy formula for the elliptic Schur functions.

For the case of trigonometric integrable models,
other boundary conditions such as the
reflecting boundaries, half-turn boundaries
are investigated \cite{Iv,BBCG,Tabony},
where in some cases symplectic Schur functions
emerge as the wavefunctions.
There are also progresses on the introduction of a
higher rank model called as the metaplectic ice \cite{BBB},
where connections with metaplectic Whittaker functions
are established. It seems valuable to lift those works
to the elliptic setting, which may lead to new developments
in number theory as well as integrable models.
\\\

\section*{Acknowledgements}
We thank the referee for giving us useful comments and letting us know
references on elliptic integrable models.
This work was partially supported by grant-in-Aid
for Scientific Research (C) No. 16K05468.

\section*{Appendix}
In this Appendix, we give a proof of \eqref{dualmaintheorem}
in some detail,
based on the idea initiated by Korepin \cite{Ko},
listing the properties of the domain wall boundary partition functions
which uniquely characterize it. This characterization lead Izergin \cite{Iz}
to found the determinant representation (Izergin-Korepin determinant) of
the domain wall boundary partition functions of the $U_q(sl_2)$
six-vertex model.
We recently extended the Izergin-Korepin technique to be able to
analyze the wavefunctions
\cite{Moelliptic,Motr,MoIK}. We apply this technique here.
As usual, we first list the properties which
uniquely characterize the wavefunctions.
\begin{proposition} 
\label{ordinarypropertiesfordomainwallboundarypartitionfunction}
The wavefunctions
$V_{M,N}(u_1,\dots,u_N|v_1,\dots,v_M|x_1,\dots,x_N|h)$
satisfies the following properties. \\
\\
 (1) If $x_N=M$, the wavefunctions
$V_{M,N}(u_1,\dots,u_N|v_1,\dots,v_M|x_1,\dots,x_N|h)$ is an elliptic polynomial of $v_M$ in $\Theta_N(\chi)$. \\
\\
 (2) The wavefunctions $V_{M,N}(u_\sigma(1),\dots,u_\sigma(N)|v_1,\dots,v_M|x_1,\dots,x_N|h)$ with the ordering of the spectral parameters permuted
$u_{\sigma(1)}, \dots, u_{\sigma(N)}$, $\sigma \in S_N$ are related to
the wavefunctions
$V_{M,N}(u_1,\dots,u_N|v_1,\dots,v_M|x_1,\dots,x_N|h)$ by
the following relation
\begin{align}
&\prod_{\substack{1 \le j < k \le N \\ \sigma(j) > \sigma(k)}}
[u_{\sigma(j)}-u_{\sigma(k)}-2p]
V_{M,N}(u_1,\dots,u_N|v_1,\dots,v_M|x_1,\dots,x_N|h) \nonumber \\
=&
\prod_{\substack{1 \le j < k \le N \\ \sigma(j) > \sigma(k)}}
[u_{\sigma(k)}-u_{\sigma(j)}-2p]
V_{M,N}(u_{\sigma(1)},\dots,u_{\sigma(N)}|v_1,\dots,v_M|x_1,\dots,x_N|h)
.
\label{permutationwavefunction}
\end{align}
\\
(3) If $x_N=M$, the following recursive relations between the
wavefunctions hold:
\begin{align}
&V_{M,N}(u_1,\dots,u_N|v_1,\dots,v_M|x_1,\dots,x_N|h)
|_{v_M=u_N+p+q_M}
\nonumber \\
=&\frac{[2p]^{1/2}[2q_M]^{1/2}[h+2\overline{q_{M-1}}]^{1/2}[h+2(N-1)p]^{1/2}}{[h+2\overline{q_M}]^{1/2}[h+2Np]^{1/2}}
\prod_{j=1}^{N-1} [u_j-u_N-2p]
\prod_{j=1}^{M-1} [u_N-v_j+p-q_j]
\nonumber \\
&\times V_{M-1,N-1}(u_1,\dots,u_{N-1}|v_1,\dots,v_{M-1}
|x_1,\dots,x_{N-1}|h)
. \label{ordinaryrecursionwavefunction}
\end{align}

If $x_N \neq M$, the following factorizations hold for the
wavefunctions:
\begin{align}
&V_{M,N}(u_1,\dots,u_N|v_1,\dots,v_M|x_1,\dots,x_N|h)
 \nonumber \\
=&\prod_{j=1}^N [u_j-v_M+q_M+p]
V_{M-1,N}(u_1,\dots,u_N|v_1,\dots,v_{M-1}|x_1
,\dots,x_N|h).
\label{ordinaryrecursionwavefunction2}
\end{align}
\\
(4) The following evaluation holds for the case $N=1$, $x_1=M$
\begin{align}
&
V_{M,1}(u|v_1,\dots,v_M|M|h) \nonumber \\
=&
\frac{[h]^{1/2}[2p]^{1/2}[2q_M]^{1/2}[u-v_M+h+p+q_M+2\overline{q_{M-1}}]}
{[h+2p]^{1/2}[h+2\overline{q_{M-1}}]^{1/2}[h+2\overline{q_{M}}]^{1/2}}
\prod_{k=1}^{M-1} [u-v_k+p-q_k].
\label{ordinaryinitialrecursion}
\end{align}
\end{proposition}

Proposition \ref{ordinarypropertiesfordomainwallboundarypartitionfunction}
can be proved in a similar way with the Proposition \ref{aboutgeneralizedscalarproducts}
for the scalar products.
The next step is to find the explicit forms of the functions
which satisfy all the properties in Proposition \ref{ordinarypropertiesfordomainwallboundarypartitionfunction}.
This is given in the next proposition.
The proof of the proposition also concludes the proof of
\eqref{dualmaintheorem}.
\begin{proposition} \label{propositionfordualwavefunction}
The product of a deformed elliptic Vandermonde determinant
and the elliptic Schur functions
$\prod_{1 \le j < k \le N} [u_j-u_k-2p]
T_{M,N}(u_1,\dots,u_N|v_1,\dots,v_M|x_1,\dots,x_N|h)$ satisfy all the properties listed in
Proposition \ref{ordinarypropertiesfordomainwallboundarypartitionfunction},
which the wavefunctions of the elliptic Felderhof model
$V_{M,N}(u_1,\dots,u_N|v_1,\dots,v_M|x_1,\dots,x_N|h)$ must satisfy.
\end{proposition}

\begin{proof}
This can be proved by showing that
\begin{align}
&H_{M,N}(u_1,\dots,u_N|v_1,\dots,v_M|x_1,\dots,x_N|h) \nonumber \\
:=
&\prod_{1 \le j < k \le N} [u_j-u_k-2p]
T_{M,N}(u_1,\dots,u_N|v_1,\dots,v_M|x_1,\dots,x_N|h) \nonumber \\
=&\prod_{1 \le j < k \le N} [u_j-u_k-2p]
\sum_{\sigma \in S_N}
\prod_{1 \le j < k \le N} \frac{1}{[u_{\sigma(j)}-u_{\sigma(k)}]}
\prod_{j=1}^N \prod_{k=x_j+1}^M [u_{\sigma(j)}-v_k+q_k+p]
\nonumber \\
&\times
\prod_{j=1}^N
\frac{[h+2(j-1)p]^{1/2} [2p]^{1/2} [2q_{x_j}]^{1/2}
}
{[h+2jp]^{1/2}[h+2\overline{q_{x_j-1}}]^{1/2}
[h+2\overline{q_{x_j}}]^{1/2}},
\nonumber \\
&\times
\prod_{j=1}^N [u_{\sigma(j)}-v_{x_j}+h+p+q_{x_j}
+2\overline{q_{x_j-1}}]
\prod_{j=1}^N \prod_{k=1}^{x_j-1}
[u_{\sigma(j)}-v_k+p-q_k],
\label{usethisforlastproperty}
\end{align}
satisfies Properties (1), (2), (3) and (4) in
Proposition \ref{ordinarypropertiesfordomainwallboundarypartitionfunction}.
For example, Property (1) can be checked by computing
the quasi-periodicities of
$H_{M,N}(u_1,\dots,u_N|v_1,\dots,v_M|x_1,\dots,x_{N-1},M|h)$ with respect
to $v_M$, which are given by
\begin{align}
&H_{M,N}(u_1,\dots,u_N|v_1,\dots,v_M+1|x_1,\dots,x_{N-1},M|h)
\nonumber \\
=&(-1)^N H_{M,N}(u_1,\dots,u_N|v_1,\dots,v_M|x_1,\dots,x_{N-1},M|h),
\label{symmetricfunctionsqpfirst} \\
&
H_{M,N}(u_1,\dots,u_N|v_1,\dots,v_M-i \mathrm{log}({\bf q})/\pi|x_1,\dots,x_{N-1},M|h)
\nonumber \\
=&(-{\bf q}^{-1})^N
\mathrm{exp} \Bigg(-2 \pi i
\Bigg(Nv_M-h-2\overline{q_{M-1}}-Nq_M-Np-\sum_{j=1}^N u_j
\Bigg)
\Bigg)
\nonumber \\
&\times H_{M,N}(u_1,\dots,u_N|v_1,\dots,v_M|x_1,\dots,
x_{N-1},M|h). \label{symmetricfunctionsqpsecond}
\end{align}
These explicit quasi-periodicities
shows that
$H_{M,N}(u_1,\dots,u_N|v_1,\dots,v_M|x_1,\dots,x_{N-1},M|h)$
is an elliptic polynomial of degree $N$ in $v_M$.
Also, the factors $(-1)^N$ and
$(-{\bf q}^{-1})^N
\mathrm{exp} \Bigg(-2 \pi i
\Bigg(Nv_M-h-2\overline{q_{M-1}}-Nq_M-Np-\sum_{j=1}^N u_j
\Bigg)
\Bigg)$ in \eqref{symmetricfunctionsqpfirst} and
\eqref{symmetricfunctionsqpsecond}
are the same with the ones which appear when we examine
the quasi-periodicities of the wavefunctions \\
$V_{M,N}(u_1,\dots,u_N|v_1,\dots,v_M|x_1,\dots,x_{N-1},M|h)$
as a function of $v_M$.

Property (3) for the case $x_N=M$ can be shown as follows.
The factors
$\displaystyle
\prod_{j=1}^{N-1} [u_{\sigma(j)}-v_M+q_M+p]
$ in each summand in
$H_{M,N}(u_1,\dots,u_N|v_1,\dots,v_M|x_1,\dots,x_N|h)$ means that we only have to deal with
the summands satisfying $\sigma(N)=N$
in \eqref{ordinaryrighthandside} which survive
after the substitution $v_M=u_N+p+q_M$.
Then we find that
$H_{M,N}(u_1,\dots,u_N|v_1,\dots,v_M|x_1,\dots,x_N|h)$
evaluated at $v_M=u_N+p+q_M$ can be expressed
using the symmetric group $S_{N-1}$
where every $\sigma^\prime \in S_{N-1}$ satisfies
$\{\sigma^\prime(1),\cdots,\sigma^\prime(N-1)\}=\{1,\cdots,N-1 \}$ as follows:
\begin{align}
&H_{M,N}(u_1,\dots,u_N|v_1,\dots,v_M|x_1,\dots,x_N|h)
|_{v_M=u_N+p+q_M} \nonumber \\
=&\prod_{1 \le j < k \le N-1} [u_j-u_k-2p]
\prod_{j=1}^{N-1} [u_j-u_N-2p] \nonumber \\
&\times \sum_{\sigma^\prime \in S_{N-1}}
\prod_{1 \le j < k \le N-1}
\frac{1}{[u_{\sigma^\prime(j)}-u_{\sigma^\prime(k)}]}
\prod_{j=1}^{N-1} \frac{1}{[u_{\sigma^\prime(j)}-u_N]} \nonumber \\
&\times \prod_{j=1}^{N-1} \prod_{k=x_j+1}^{M-1}
[u_{\sigma^\prime(j)}-v_k+q_k+p]
\prod_{j=1}^{N-1} [u_{\sigma^\prime(j)}-u_N] \nonumber \\
&\times \frac{[h+2(N-1)p]^{1/2}[2p]^{1/2}[2q_M]^{1/2}}
{[h+2Np]^{1/2}[h+2\overline{q_{M-1}}]^{1/2}[h+2\overline{q_M}]^{1/2}}
\prod_{j=1}^{N-1}
\frac{[h+2(j-1)p]^{1/2} [2p]^{1/2} [2q_{x_j}]^{1/2}
}
{[h+2jp]^{1/2}[h+2\overline{q_{x_j-1}}]^{1/2}
[h+2\overline{q_{x_j}}]^{1/2}}
\nonumber \\
&\times
[h+2\overline{q_{M-1}}]
\prod_{j=1}^{N-1}
[u_{\sigma^\prime(j)}-v_{x_j}+h+p+q_{x_j}
+2\overline{q_{x_j-1}}]
\nonumber \\
&\times
\prod_{j=1}^{N-1} \prod_{k=1}^{x_j-1}
[u_{\sigma^\prime(j)}-v_k+p-q_k]
\prod_{k=1}^{M-1}
[u_N-v_k+p-q_k].
\label{ordinaryrighthandsideaftersubstitution}
\end{align}
After appropriately cancelling and rearraning the expression above,
we find that \eqref{ordinaryrighthandsideaftersubstitution}
can be rewritten as
\begin{align}
&H_{M,N}(u_1,\dots,u_N|v_1,\dots,v_M|x_1,\dots,x_N|h)
|_{v_M=u_N+p+q_M}
\nonumber \\
=&\frac{[2p]^{1/2}[2q_M]^{1/2}[h+2\overline{q_{M-1}}]^{1/2}[h+2(N-1)p]^{1/2}}{[h+2\overline{q_M}]^{1/2}[h+2Np]^{1/2}}
\prod_{j=1}^{N-1} [u_j-u_N-2p]
\prod_{j=1}^{M-1} [u_N-v_j+p-q_j]
\nonumber \\
&\times \prod_{1 \le j < k \le N-1} [u_j-u_k-2p]
\sum_{\sigma^\prime \in S_{N-1}}
\prod_{1 \le j < k \le N-1}
\frac{1}{[u_{\sigma^\prime(j)}-u_{\sigma^\prime(k)}]} \nonumber \\
&\times
\prod_{j=1}^{N-1} \prod_{k=x_j+1}^{M-1}
[u_{\sigma^\prime(j)}-v_k+q_k+p]
\prod_{j=1}^{N-1}
\frac{[h+2(j-1)p]^{1/2} [2p]^{1/2} [2q_{x_j}]^{1/2}
}
{[h+2jp]^{1/2}[h+2\overline{q_{x_j-1}}]^{1/2}
[h+2\overline{q_{x_j}}]^{1/2}}
\nonumber \\
&\times
\prod_{j=1}^{N-1}
[u_{\sigma^\prime(j)}-v_{x_j}+h+p+q_{x_j}
+2\overline{q_{x_j-1}}]
\prod_{j=1}^{N-1} \prod_{k=1}^{x_j-1}
[u_{\sigma^\prime(j)}-v_k+p-q_k]
\nonumber \\
=&\frac{[2p]^{1/2}[2q_M]^{1/2}[h+2\overline{q_{M-1}}]^{1/2}[h+2(N-1)p]^{1/2}}{[h+2\overline{q_M}]^{1/2}[h+2Np]^{1/2}}
\prod_{j=1}^{N-1} [u_j-u_N-2p]
\prod_{j=1}^{M-1} [u_N-v_j+p-q_j]
\nonumber \\
&\times H_{M-1,N-1}(u_1,\dots,u_{N-1}|v_1,\dots,v_{M-1}
|x_1,\dots,x_{N-1}|h),
\end{align}
which is exactly the same with the one
\eqref{ordinaryrecursionwavefunction} for the wavefunctions
of the elliptic Felderhof model
$V_{M,N}(u_1,\dots,u_N|v_1,\dots,v_M|x_1,\dots,x_N|h)$.

\end{proof}

Now that we have shown
Proposition \ref{propositionfordualwavefunction},
we find that the elliptic function \\
$H_{M,N}(u_1,\dots,u_N|v_1,\dots,v_M|x_1,\dots,x_N|h)$ is nothing but the explicit form of the wavefunctions
$V_{M,N}(u_1,\dots,u_N|v_1,\dots,v_M|x_1,\dots,x_N|h)$
\begin{align}
V_{M,N}(u_1,\dots,u_N|v_1,\dots,v_M|x_1,\dots,x_N|h)
=
H_{M,N}(u_1,\dots,u_N|v_1,\dots,v_M|x_1,\dots,x_N|h),
\end{align}
hence \eqref{dualmaintheorem} in Theorem \ref{maintheoremstatement}
is proved.

\bibliographystyle{plainnat}

\end{document}